%Paper: hep-th/9503036
%From: J.Park@swansea.ac.uk
%Date: Tue, 07 Mar 95 03:22:11 +0000
%Date (revised): Fri, 17 Mar 95 10:59:15 +0000

%This is THE FINAL VERSION

%------------------------------------------------------------------
%You need AMSfonts 2.1. If you do not have
%AMSfonts, please search for the word "Euler" and follow the instructions.
%------------------------------------------------------------------

%\input lanlmac
\input harvmac
\def\bigans{y }
\message{ Do you want the figures  (y/n)? }\read-1 to\answ
\ifx\answ\bigans\message{(This will come out with figures.}
\input epsf
\else\message{(No figures.}
\fi

%------------------------------------------------------------------
% personal macros
%------------------------------------------------------------------
\def\title#1#2#3#4{\nopagenumbers
\abstractfont\hsize=\hstitle
\rightline{YUMS-#1,
SWAT/#2, hep-th/9503036}%
\bigskip
\vskip 0.7in\centerline{\titlefont #3}\abstractfont\vskip .2in
\centerline{\titlefont #4}\abstractfont\vskip .2in
\vskip 0.3in\pageno=0
}

%\draftmode
%\baselineskip=16pt plus 2pt minus 1pt
\newskip\normalparskip
\normalparskip = 6pt plus 2pt minus 1pt
\parskip = \normalparskip
\parindent=12pt

%------------------------------------------------------------------
%symbols
%------------------------------------------------------------------
\def\a{\alpha}           \def\c{\chi}       \def\d{\delta}
    \def\e{\varepsilon}        
    \def\G{\Gamma}           
\def\L{\Lambda}   \def\m{\mu}                 
  \def\o{\omega}      \def\O{\Omega}     \def\p{\psi}
      \def\s{\sigma}      \def\S{\Sigma}     \def\th{\theta}
        \def\w{\varphi}    

\def\CA{{\cal A}}

\def\CC{{\cal C}}
\def\CG{{\cal G}}
\def\CM{{\cal M}}
\def\CN{{\cal N}}
\def\CE{{\cal E}}
\def\CL{{\cal L}}
\def\CD{{\cal D}}
\def\CW{{\cal W}}

%
%------------------------------------------------
% Euler Fonts
\font\teneufm=eufm10
\font\seveneufm=eufm7
\font\fiveeufm=eufm5
\newfam\eufmfam
\textfont\eufmfam=\teneufm
\scriptfont\eufmfam=\seveneufm
\scriptscriptfont\eufmfam=\fiveeufm
\def\eufm#1{{\fam\eufmfam\relax#1}}

\font\teneusm=eusm10
\font\seveneusm=eusm7
\font\fiveeusm=eusm5
\newfam\eusmfam
\textfont\eusmfam=\teneusm
\scriptfont\eusmfam=\seveneusm
\scriptscriptfont\eusmfam=\fiveeusm

\font\tenmsx=msam10
\font\sevenmsx=msam7
\font\fivemsx=msam5
\font\tenmsy=msbm10
\font\sevenmsy=msbm7
\font\fivemsy=msbm5
\newfam\msafam
\newfam\msbfam
\textfont\msafam=\tenmsx  \scriptfont\msafam=\sevenmsx
  \scriptscriptfont\msafam=\fivemsx
\textfont\msbfam=\tenmsy  \scriptfont\msbfam=\sevenmsy
  \scriptscriptfont\msbfam=\fivemsy

\def\msbm#1{{\fam\msbfam\relax#1}}

%\def\eufm#1{{\bf #1}}  %If you do not have AMSfonts 2.1,
%\def\eusm#1{{\cal #1}} %uncomment these three lines
%\def\msbm#1{{\bf #1}}  %and comment the lines below %Euler Fonts

%-----------------------------------------------------------------
%Math
\def\rd{\partial}

\def\darr#1{\raise1.5ex\hbox{$\leftrightarrow$}\mkern-16.5mu #1}
\def\Ha{{1\over2}}

\def\fr#1#2{{\textstyle{#1\over#2}}}
\def\Fr#1#2{{#1\over#2}}
\def\tr{\hbox{Tr}\,}

   %Feynman dagger
 %Feynman dagger
\def\roughly#1{\raise.3ex\hbox{$#1$\kern-.75em\lower1ex\hbox{$\sim$}}}

%
%Journals
\def\cmp#1#2#3{Commun.\ Math.\ Phys.\ {{\bf #1}} {(#2)} {#3}}
\def\pl#1#2#3{Phys.\ Lett.\ {{\bf #1}} {(#2)} {#3}}
\def\np#1#2#3{Nucl.\ Phys.\ {{\bf #1}} {(#2)} {#3}}

\def\ijmp#1#2#3{Int.\ J.\ Mod.\ Phys.\ {{\bf #1}} {(#2)} {#3}}

\def\jdg#1#2#3{J.\ Differ.\ Geom.\ {{\bf #1}} {(#2)} {#3}}

\def\top#1#2#3{Topology {{\bf #1}} {(#2)} {#3}}
\def\zp#1#2#3{Z.\ Phys.\ {{\bf #1}} {(#2)} {#3}}
\def\prp#1#2#3{Phys.\ Rep.\ {{\bf #1}} {(#2)} {#3}}

\def\ptrsls#1#2#3{Philos.\ Trans.\  Roy.\ Soc.\ London
{{\bf #1}} {(#2)} {#3}}

\def\im#1#2#3{Invent.\ Math.\ {{\bf #1}} {(#2)} {#3}}
\def\plms#1#2#3{Proc.\ London Math.\ Soc.\ {{\bf #1}} {(#2)} {#3}}
\def\dmj#1#2#3{Duke Math.\  J.\ {{\bf #1}} {(#2)} {#3}}
\def\bams#1#2#3{Bull.\ Am.\ Math.\ Soc.\ {{\bf #1}} {(#2)} {#3}}
\def\jgp#1#2#3{J.\ Geom.\ Phys.\ {{\bf #1}} {(#2)} {#3}}
%------------------------------------------------------------------
%Macros specific to this file

\def\pr{\prime}
\def\ppr{{\prime\prime}}

\def\bs{{\bf s}}
\def\bbs{{\bf\bar s}}
\def\Da{d_{\!A}}
\def\Dp{\rd_{\!A}}

\def\Dpp{\bar\rd_{\!A}}

\def\gE{\eufm{g}_{\raise-.1ex\hbox{${}_E$}}}
\def\gBE{\eufm{g}_{\raise-.1ex\hbox{${}_\BE$}}}
\def\gEc{\eufm{g}_{\raise-.1ex\hbox{${}_E$}}^C}
\def\BC{\msbm{C}}
\def\BE{\msbm{E}}

\def\BR{\msbm{R}}
\def\BZ{\msbm{Z}}

\def\lin#1{\medskip\leftline{$\underline{\hbox{\it #1}}$}\medskip}

%------------------------------------------------------------------
\lref\Atiyah{
M.F.~Atiyah,
Convexity and commuting Hamiltonians,
Bull.~Lond.~Math.~Soc.~{\bf 14} (1982) 1.
}
\lref\ABa{
M.F.\ Atiyah and R.\ Bott,
The Yang-Mills equations over Riemann surfaces,
\ptrsls{A 308}{1982}{523}.
}
\lref\ABb{
M.F.~Atiyah and R.~Bott,
The moment map and equivariant cohomology,
\top{23}{1984}{1}.
}
\lref\AJ{
M.F.~Atiyah and L.~Jeffrey,
Topological Lagrangians and cohomology,
\jgp{7}{1990}{1}.
}
\lref\Blau{
M.~Blau,
The Mathai-Quillen formalism and topological field theory,
\jgp{11}{1991}{129}.
}
\lref\BTHa{
M.~Blau and G.~Thompson,
 Lectures on 2d Gauge Theories: Topological Aspects and Path Integral
 Techniques, preprint  hep-th/9310144.
 }
\lref\BTHb{
M.~Blau and G.~Thompson,
On Diagonalization in Map(M,G),
preprint hep-th/9402097.
}
\lref\BTHc{
M.~Blau and G.~Thompson,
Localization and Diagonalization: A review of functional integral
techniques for low-dimensional gauge theories and topological field theories,
preprint hep-th/9501075
}
\lref\BMS{
R.\ Brooks, D.\ Montano and J.\ Sonnenschein,
Gauge fixing and renormalization in topological quantum field theory.
\pl{B 214}{1988}{91}
}
\lref\BRT{
D.\ Birmingham, M.\ Rakowski and G.\ Thompson,
\np{B 315}{1989}{577}
}
\lref\BS{
L.\ Baulieu and I.M.\ Singer,
Topological Yang-Mills symmetry,
\np{(Proc.\ Suppl.) 5B}{1988}{12}.
}
\lref\BGV{
N.~Berline, E.~Getzler and M.~Vergne, Heat kernels and Dirac Operators
(Springer-Verlag,1992).
}
\lref\BT{
R.~Bott and L.~Tu, Differential forms in algebraic topology (Springer-Verlag,
1982).
}
\lref\CMR{
S.~Cordes, G.~Moore and S.~Ramgoolam,
Lectures on 2D Yang-Mills theory, equivariant cohomology and topological field
theories, Part II, preprint hep-th/9411210.
}
\lref\DonaldsonA{
S.K.\ Donaldson,
Anti-self-dual Yang-Mills connections
on complex algebraic surfaces and stable bundles,
\plms{30}{1985}{1}\semi
Infinite determinants, stable bundles and curvature,
\dmj{54}{1987}{231}.
}
\lref\DonaldsonB{
S.K.~Donaldson,
Connections, cohomology and the intersection forms of four manifolds,
\jdg{24}{1986}{275}.
}
\lref\DonaldsonC{
S.K.\ Donaldson,
Irrationality and the $h$-cobordism conjecture.
\jdg{26}{1987}{141}.
}
\lref\DonaldsonD{
Donaldson, S.K.:
The orientation of Yang-Mills moduli spaces and $4$-manifold topology.
\jdg{26}{1987}{397}.
}
\lref\DonaldsonE{
S.K.\ Donaldson, Polynomial invariants for smooth $4$-manifolds,
\top{29}{1990}{257}.
}
\lref\DonaldsonF{
S.K.~Donaldson, Yang-Mills Invariants of four-manifolds.
In: Geometry of low-dimensional manifolds $1$, Donaldson, S.K.,
Thomas, C.B. (eds.),London Mathematical Society Lecture Note Series
$150$  ( Cambridge Univ.\ Press 1990).
}
\lref\DH{
J.J.\ Duistermaat and G.J.\ Heckmann,
On the variation in the cohomology
in the symplectic form of the reduced phase space,
Invent Math.\ {\bf 69} (1982) 259.
}
\lref\DoH{
I.V.~Dolgachev and Y.~Hu,
Variation of geometrical invariant theory Quotients,
preprint.
}
\lref\DK{
S.K.\ Donaldson and P.B.\ Kronheimer,
The geometry of four-manifolds (Oxford University Press,  1990).
}
\lref\EG{
G.~Ellingsrud and L.~G\"{o}ttsche,
Variation of moduli spaces and Donaldson invariant under change of
polarization,
preprint alg-geom/9410005 (the expanded version).
}
\lref\FM{
R.~Friedman and J.W.~Morgan,
On the diffeomorphism types of certain algebraic surfaces I, II,
\jdg{27}{1988}{297}.
}
\lref\FQA{
R.~Friedman and Z.~Qin,
On complex surfaces diffeomorphic to rational surfaces,
preprint alg-geom/9406
}
\lref\FQB{
R.~Friedman and Z.~Qin,
Flips of moduli spaces and transition formulas for Donaldson polynomial
invariants of rational surfaces,
preprint alg-geom/9410007.
}
\lref\GO{
A.\ Galperin and O.\ Ogievetsky,
Holonomy groups, complex structures and $D=4$ topological Yang-Mills
theory,
\cmp{139}{1991}{377}.
}
\lref\GS{
V.~Guillemin and S.~Sternberg,
Birational equivalence in symplectic category,
\im{97}{1989}{485}.
}
\lref\Ger{
A.~Gerasimov,
Localization in GWZW and Verlinde formula,
preprint hep-th/9305090.
}
\lref\JK{
L.C.~Jeffrey and F.C.~Kirwan,
Localization for non-abelian group actions,
preprint alg-geom/9307005.
}
\lref\KalkA{
J.~Kalkman,
BRST model for equivariant cohomology and representatives for the
equivariant Thom class,
\cmp{153}{1993}{447}.
}
\lref\KalkB{
J.~Kalkman,
Cohomology rings of symplectic quotients,
J.~Reine Angew.~ Math., to appear.
}
\lref\KalkC{
J.~Kalkman,
Residues in nonabelian localization,
preprint hep-th/9407168.
}
\lref\Kanno{
H.\ Kanno,
Weil algebraic structure and geometrical meaning of the BRST
transformation in topological quantum field theory,
\zp{C 43}{1989}{477}.
}
\lref\Kirwan{F.\ Kirwan, Cohomology of quotients in symplectic
and algebraic geometry (Princeton Univ.~Press, 1987).
}
\lref\KotsA{
D.~Kotschick,
On manifolds homeomorphic to $CP^2\# 8\overline{CP}^2$,
\im{95}{1989}{591}.
}
\lref\KotsB{
D.~Kotschick,
$SO(3)$-invariants for $4$-manifolds with $b^+_2 =1$,
\plms{63}{1991}{426}.
}
\lref\KoM{
D.~Kotschick and J.W.~Morgan,
$SO(3)$-invariants for $4$-manifolds with $b^+_2 =1$.~II,
\jdg{39}{1994}{433}.
}
\lref\KM{
P.~Kronheimer and T.~Mrowka, Recurrence relations and asymptotics
for four-manifold invariants, \bams{30}{1994}{215}.
}
\lref\KMb{
P.~Kronheimer and T.~Mrowka,
The genus of embedded surfaces in the projective plane,
preprint, 1994.
}
\lref\Hitchin{
N.~Hitchin,
The geometry and topology of moduli space,
in Global geometry and Mathematical Physics, eds. L.~Alvarez-Gaum\'{e}
et.~als., LNM 1451 (Springer-Verlag, 1992).
}
\lref\HL{
Y.~Hu and W.-P.~Li,
Variation of the Gieseker and Uhlenbeck compactifications,
preprint alg-geom/9409003.
}
\lref\HPa{
S.J.~Hyun and J.-S.~Park,
$N=2$ Topological Yang-Mills Theories and
Donaldson's polynomials, preprint hep-th/9404009[revised by Aug., 1994],
submitted to J.~Geom.~Phys.
}
\lref\HPb{
S.J.~Hyun and J.-S.~Park,
The $N=2$ supersymmetric quantum field theories and the Dolbeault
model of the equivariant cohomology,  in preparation.
}
\lref\HPP{
S.J.~Hyun, J.~Park and J.-S.~Park,
Topological QCD, to appear.
}
\lref\Li{J.~Li,
Algebraic geometric interpretation of Donaldson's polynomial invariants,
\jdg{37}{1993}{417}.
}
\lref\LP{
J.M.F.\ Labastida and M.\ Pernici,
A gauge invariant action in topological quantum field theory.
\pl{B 212}{1988}{56}
}
\lref\Mong{K.C.~Mong
Polynomial invariants for $4$-manifolds of type $(1,n)$
and a calculation for $S^2\times S^2$,
Quart.~J.~Math.~Oxford,{\bf 43} (1992) 459.
}
\lref\Morgan{J.W.~Morgan,
Comparison of the Donaldson polynomial invariants with their algebro-geometric
analogues,
\top{32}{1993}{449}.
}
\lref\MorganB{J.W.~Morgan,
Lecture given at the Newton Inst., Dec.~1994.
}
\lref\MQ{
V.~Mathai and D.~Quillen,
Thom classes, superconnections and equivariant differential forms,
\top{25}{1986}{85}.
}
\lref\MW{
K.~Matsuki and R.~Wentworth,
Mumford-Thaddeus principle on the moduli space of vector bundles on an
algebraic surface,
preprint alg-geom/9410016.
}
\lref\OSB{
S.\ Ouvry, R.\ Stora and P.\ van Ball,
Algebraic characterization of topological Yang-Mills theory,
\pl{B 220}{1989}{1590}.
}
\lref\ParkA{
J.-S.\ Park,
$N=2$ topological Yang-Mills theory on compact
K\"{a}hler surfaces, \cmp{163}{1994}{113}.
}
\lref\ParkB{
J.-S.\ Park, Holomorphic Yang-Mills theory on compact
K\"{a}hler Manifolds, \np{B423}{1994}{559}.
}
\lref\PW{
E.~Prato and S.~Wu,
Duistermaat-Heckman measures in a non-compact setting,
preprint alg-geom/9307005.
}
\lref\QinA{
Z.~Qin, Complex structures on certain differentiable $4$-manifolds,
\top{32}{1993}{551};
Equivalence classes of polarizations and moduli spaces of sheaves,
\jdg{37}{1993}{397}.
}
\lref\LQ{
W.-P.~Li and Z.~Qin, Low-degree Donaldson polynomial invariants
of rational surfaces, J.~Alg.~Geom.~{\bf 37} (1993) 417.
}
\lref\REPORT{
D.\ Birminham, M.\ Blau, M.\ Rakowski and G.\ Thomson,
%Topological field theory.
\prp{209}{1991}{129}.
}
\lref\ThA{
M.~Thaddeus,
Stable pairs, linear systems and the Verlinde formula,
preprint alg-geom/9210007.
}
\lref\ThB{
M.~Thaddeus,
Geometric invariant theory and flips,
preprint alg-geom/9405004 .
}
%\lref\ThC{
%M.~Thaddeus,
%Toric quotients and flips,
%to appear in {\it Proceedings of the 1993 Taniguchi symposium}.
%}
\lref\Thompson{
G.~Thompson,
1992 Trieste Lectures on Topological Gauge Theory and Yang-Mills Theory,
}
\lref\SWa{
N.~Seiberg and E.~Witten,
Electric-magnetic duality, monopole condensation, and confinement
in $N=2$ supersymmetric Yang-Mills theory,
\np{B 426}{1994}{19}.
}
\lref\SWb{
N.~Seiberg and E.~Witten,
Monopoles, duality, and chiral symmetry breaking in $N=2$ supersymmetric
QCD,
\np{B 431}{1994}{484}.
}
\lref\Verg{
M.~Vergne,
A note on Jeffrey-Kirwan-Witten's localization formula, preprint LMENS -94-12.
}
\lref\VW{
C.~Vafa and E.~Witten,
A strong coupling test of $S$-duality,
\np{B 431}{1994}{3}
}
\lref\WittenA{
E.~Witten,
Topological quantum field theory,
\cmp{117}{1988}{353}.
}
\lref\WittenB{
E.~Witten,
Supersymmetric Yang-Mills theory on a four manifolds,
J.~Math.~Phys.~{\bf 35} (1994) 5101.
%Preprint IASSNS-HEP-94/5 February, 1994.
%hep-th/9403195.
}
\lref\WittenC{
E.~Witten,
Two dimensional gauge theories revisited.
\jgp{9}{1992}{303}.
}
\lref\WittenD{
E.~Witten,
The $N$ matrix model and gauged WZW models,
\np {B 371}{1992}{191}; Mirror manifolds and topological field theory,
in {\it Essays on mirror manifolds,}
ed.~S.-T.~Yau (International Press, 1992).
}
\lref\WittenE{
E.~Witten,
Monopoles and four-manifolds,
preprint  hep-th/9411102.
}
\lref\WittenF{
E.~Witten,
Phase of $N=2$ models in two dimensions,
\np{B 403}{1993}{159}.
}
\lref\WittenG{
E.~Witten,
Supersymmetry and Morse theory,
\jdg{16}{1982}{353}.
}
\lref\WittenAA{
E. Witten,
Introduction to cohomological field theories.
\ijmp{A 6}{1991}{2273}.
}
\lref\Wu{
S.~Wu,
An integration formula for the square of moment maps of circle actions,
preprint hep-th/9212071.
}
\lref\FA{
D.~Anselmi and P.~Fre',
Gauged Hyperinstantons and Monopole Equations,
preprint  hep-th/9411205 and references therein.
}

%--------------------------------------------------------------
\title{94-23}{65}{Holomorphic Yang-Mills Theory}
{ and Variation of the Donaldson  Invariants}

\centerline{
Seungjoon Hyun\footnote{$^{\dagger}$}{e-mail: hyun@phya.yonsei.ac.kr}
}
\bigskip
\centerline{{\it Institute for Mathematical Sciences}}
\centerline{{\it Yonsei University}}
\centerline{{\it Seoul 120-749, Korea}}
\bigskip
\bigskip
\centerline{
Jae-Suk Park\footnote{$^{\dagger\dagger}$}{e-mail:
j.park@swansea.ac.uk}}
\bigskip
\centerline{\it Theoretical Physics Group}
\centerline{\it Department of Physics}
\centerline{\it University of Wales, Swansea}
\centerline{\it Swansea SA2 8PP, UK }
\bigskip
\bigskip

%abstract
\centerline{
{\bf Abstract}
}
\vskip 0.2in
We study the path integrals of the holomorphic Yang-Mills theory
on compact K\"{a}hler surface with $b_2^+ = 1$.
Based on the results, we examine  the correlation functions of the topological
Yang-Mills theory and the corresponding  Donaldson invariants
as well as their transition formulas.

\Date{Feb, 1995}

%\submit

\newsec{Introduction}
The Donaldson polynomial invariants are powerful tools for classifying the
smooth structures of four-manifolds \DonaldsonE\DK.  For a Riemann manifold
with
$b^{+}_2 \geq 3$, these polynomials are  well-defined  and metric
independent, as long as it is generic.

On the other hand, for a  manifold with $b^+_2=1$,  the polynomials depend on
metric in a very interesting way. This is due to the reducible anti-self-dual
(ASD) connections which appear at finite points of the generic smooth path
connecting two generic metrics. The  first  example was studied by Donaldson in
his seminal paper on the failure of the h-cobordism conjecture
\DonaldsonC.
His formula was further studied  in detail by Friedman-Morgan \FM, and
generalized
by Kotschick \KotsB, Mong  \Mong, and Kotschick-Morgan \KoM.

The topological Yang-Mills  (TYM) theory  proposed by Witten is a field
theoretic interpretation of the Donaldson invariants \WittenA.
The relation with  the theory to the BRST quantization were much studied in
\BS\LP\BMS\BRT\Kanno\OSB.\foot{We  refer to ref. \REPORT\
as a review and for further references. }
It is also a new mathematical viewpoint on the Donaldson invariants
due to the Atiyah-Jeffrey re-interpretation \AJ\ of the
Witten's approach as an infinite dimensional generalization
of the Mathai-Quillen representative \MQ\ of the equivariant Thom
class.
The TYM theory on compact K\"{a}hler surface was studied in  some detail
by the second  author \ParkA. It was shown that the theory has two global
fermionic
symmetries.\foot{The first  formulation of the TYM theory on K\"{a}hler surface
was given by \GO.}
The theory can be interpreted as an infinite dimensional
version of a Dolbeault equivariant cohomological analogue of the
Mathai-Quillen formalism \HPa\HPb.
Witten studied the K\"{a}hler case by twisting the $N=2$ super-Yang-Mills
theory
and determined the Donaldson Polynomial invariants for K\"{a}hler surfaces with
$b^+_2\geq 3$ almost completely\WittenB. His approach gives yet another new
perspective on the Donaldson theory since he related the theory to the infrared
behavior of the $N=1$ super-Yang-Mills theory.

On the other hand, to the authors' knowledge, no serious attempt has been  made
for the field theoretic study of the Donaldson polynomial  invariants for
manifolds  with $b^+_2=1$.\foot{In ref.~\VW, it was shown that the variation of
the Donaldson invariants is related to the holomorphic anomaly.}
The TYM theory may  not be well-defined  in the presence of the reducible
anti-self-dual (ASD)  connections.  The reducible connections also contribute
to
the
non-compactness of the space where  the path integral is  localized,
due to the flat directions of the scalar potential.
This makes the topological interpretation of the theory
unclear \WittenAA.
Furthermore, the crucial perturbation of Witten \WittenB, inducing the mass gap
to the theory, is not applicable to this case\foot{In ref.~\HPa, we have
exploited the origin of the mass gap in terms of the Dolbeault model of the
equivariant cohomology.}.

In this paper, we study the $SU(2)$ Donaldson polynomial
invariants of  a simply connected compact K\"{a}hler surface
$X$ with the vanishing geometric genus  $p_g(X) = 0$,
i.e., $b^{+}_2(X) = 1$, based on the
holomorphic Yang-Mills (HYM) theory \ParkB.
The HYM theory was proposed by the second author
to provide yet another field theoretic interpretation of the Donaldson
invariants of K\"{a}hler surface,  adopting the two dimensional construction of
 Witten \WittenC. It was shown that there exists a simple mapping to TYM theory
on K\"{a}hler surface analoguos to the mapping from two-dimensional TYM theory
to the physical YM theory.
It turns out that the HYM theory is more useful
for manifold with $p_g =0$ \foot{The HYM theory was also used
to determine the non-algebraic part of the  Donaldson invariants of manifold
with $H^{2,0} \neq 0$ \HPa.}. A la HYM theory we will show that the path
integral approach to the Donaldson theory is well-defined whatever properties
the moduli space of ASD connections has \WittenA.

The basic idea of the HYM theory is that the moduli space of ASD connections
over K\"{a}hler surface is the symplectic quotients in the space $\CA^{1,1}$
of all holomorphic connections. Classically, the action functional of the
HYM theory is the norm squared of moment map.  The partition function
of the theory   can be expressed by the values of
the moment map at the critical points of
the action functional. The spectrum of the  critical
points, as one varies
the metric, depends on a certain chamber structure of the positive
cone. By studying the partition function and some correlation functions in the
small coupling limit, one can obtain the expectation values of certain
topological observables of TYM theory.  It turns out those expectation
values are well-defined and depend only on the chamber structure of
the metric.

The basic method of our calculation is the fixed point theorem of
Witten \WittenD. Unfortunately, we could calculate only
one of the two branches of the fixed points. The branch we calculate
does not contain any information on the genuine diffeomorphism invariants.
While we were struggling with this problem, the Donaldson theory evolved
into the new dimension by the fundamental works of Seiberg and Witten
on the strong coupling properties of the $N=2$ super-Yang-Mills theory
\SWa\SWb. By exploiting a dual Donaldson theory (Seiberg-Witten theory),
Witten determined the invariants completely for K\"{a}hler surface
with $b_2^+ \geq 3$\WittenE. This new method resolves all of the difficulties
of
the
Donaldson theory as well as the  TYM  theory, such as the non-compactness of
the
moduli space of ASD connections. For a manifold with $b_2^+ =1$,
Witten's new approach has an additional complication comparing
to the case with $b_2^+\geq 3$\WittenE.
The calculation we have done
in this paper can be viewed as the path integral contributed from
the generic points of the quantum moduli space (the complex $u$-plane in
\WittenE)
except the two singular points.
We will argue that the remaining branch of the partition function of HYM
theory may be obtained by the Seiberg-Witten invariants which
corresponds to the path integral contributed from the
two singular points in the quantum moduli space.

As a matter of fact, the  expectation values of TYM theory are not identical
with the corresponding Donaldson invariants.
The difference is originated from  the compactification of the moduli space
ASD connections  in the mathematical definition of the Donaldson invariants.
In the field theoretic approach, both TYM and HYM theories
do not refer to any compactification procedure. In any case, the path
integrals are turned out to be  well-defined though the moduli space is
rarely compact.  To recover the original Donaldson invariants
some explicit prescription for incorporation of the compactification
may be required. There are many reasons that the only
difference between the two approaches is that the Donaldson
theory has additional  contributions due to
reducible connections with lower instanton numbers.
However, the additional part does not have any relevant information
on the diffeomorphism class of the underlying manifolds.

We will interpret our results as the contributions
of the original moduli space of ASD connections which is the top
stratum of the compactified moduli space.
Then, our results can be used to determine the transition formula
of the Donaldson invariants.   In some respects our results are
more general than the known  mathematical results.
We also determine the non-analytic (non-polynomial) property of the invariants
 when the moduli space is singular. We  believe that our
formula gives a very clear and an intuitive understanding
of the mechanism of  the variation of the Donaldson invariants.
This allows us to predict a rather detailed formula of the Donaldson
invariants if the transition formula is known.

The invariants and the  explicit  transition formulas
for some special cases were  calculated in  \QinA\LQ.
While we are preparing this  paper the general transition formula appeared in
term of some enumerative geometry of the Hilbert schemes\FQB\EG\MW.
The paper of Friedman and Qin \FQB\ contains some explicit results on the
transition formula.  The  paper of Ellingsrud and G\"{o}ttsche \EG
contains more general explicit results.\foot{ We would like to thank
L.~G\"{o}ttsche for pointing us out, after our first version,  the expanded
version of \EG\ which contains the explicit results.}.
Their mathematical picture on the variation of the moduli space
is quite similar to our physical picture.  At the
moment, the invariants are calculated  only for some special cases and  no
general explicit  transition formula  is known \QinA\LQ.

This paper is organized as follows.
In Sect.~2, we review the
Donaldson polynomial invariants, the TYM theory and the HYM theory.
We show that the HYM theory is a natural field theoretic model
for the Donaldson invariants of manifolds with $p_g=0$.
In Sect.~3,  we give the path integral calculation of the
HYM theory and obtain an effective theory.
In Sect.~4, we examine the small coupling behavior of the partition
function and expectation values of topological observables.
Then, we study the  variation  of the path integral according to
the variation of the metric.
In Sect.~5, we obtain the expectation values in the TYM theory
from the result of HYM theory. We argue that the unknown branch
of the path integral corresponds to the Seiberg-Witten monopole invariants.
After a brief discussion of some properties of the monopole invariants,
we show that the expectation values of TYM theory depends only
on certain chamber structures in the space of metric.
In Sect.~6, we discuss  some natural implications of our results on
the Donaldson invariant and  its transition formula, as well as
some conjectures and  speculations.

\newsec{The Donaldson Polynomials and  Holomorphic Yang-Mills Theory}

In this section, we review the Donaldson theory, TYM theory and HYM theory
on  a manifold with $b^+_2(X) = 1$.  Throughout this paper, for simplicity,
we consider a simply connected compact algebraic surface $X$ and the $SU(2)$
polynomial invariants only.
There is no loss of generality to consider the algebraic surface only
since any K\"{a}hler surface deforms to an algebraic one.
We hope we can treat more general gauge group in the future.

Let $X$ be a simply connected projective algebraic surface with
K\"{a}hler-Hodge
metric $g$ and associated K\"{a}hler form $\o$ of type $(1,1)$.
Since $X$ is a projective space,
we have an ample line bundle $H$ over $X$ whose first Chern Class
$c_1(H) \in H^2(X;\BZ)$ is given by the K\"{a}hler form $\o$.
We will occasionally confuse  a line bundle with its first Chern class
as well as with the the corresponding divisor.
By the Kodaria projective embedding theorem,  sections of some positive
tensor power $H^{\otimes m}$ of the ample line bundle define projective
embedding of $X$ to some complex projective space whose hyperplane
section class is Poincar\'{e} dual to $m[\o]$.
A line bundle $L$ is ample if and only if $L\cdot L > 0$ and $L\cdot c > 0$
for every complex curve $c$ in $X$, where $`\cdot `$ denotes
the intersection pairings, i.e., $L\cdot L \equiv \int_X c_1(L)\wedge c_1(L)$.
They defines a positive cone called ample cone (or K\"{a}hler cone)
$Cone_X$ in the space of  $H^{2}(X;\BR)$.

For a given metric
we can decompose the space $H^2(X)$ of harmonic forms on $X$ into the
space  $H_+^2(X)$ of self-dual and the space $H_-^2(X)$ of anti-self-dual
harmonic forms. Similarly,  we have $b_2 = b_2^+ + b_2^-$ where
 $b_2 = \hbox{dim}\, H^2(X)$ and $b_2^\pm = \hbox{dim}\,H_\pm^2(X)$.
 From the Hodge index  theorem, we have
\eqn\aaa{
b_2^+ = 1 + 2p_g,
}
where the one-dimensional piece is spanned by the K\"{a}hler form $\o$ and the
geometric genus $p_g$ is the number of (harmonic) holomorphic two-forms.
For an arbitrary vector bundle valued two-form we have the similar
decompositions, i.e.~the self-dual part of the curvature two-form $F^+(A)$
can be decomposed as
\eqn\aaaa{
F^+(A) = F^{2,0}(A) + f(A) \o + F^{0,2}(A),
}
where $f(A) = \Ha \L F^{1,1}(A)$. Here $\L$ denotes the algebraic trace
operator
which is adjoint to the wedge multiplication of the K\"{a}hler form.
The intersection form $q_X$ between harmonic two-forms is an unimodular
(due to the Poincar\'{e} duality) and  symmetric $b_2\times b_2$ matrix
with signature $(b_2^+ - b_2^-)$.
That is, upon diagonalization of $q_X$ it has $b_2^+$ positive entries and
$b_2^-$ negative entries. It is also called of type $(b_2^+,b_2^-)$.
This comes from the simple fact that
\eqn\aab{\eqalign{
q_X(\a, \a) = \a\cdot \a = \int_X \a\wedge\a
&= \int_X \a^+\wedge\a^+ + \int_X \a^-\wedge\a^-\cr
&=\int_X \a^+\wedge *\a^+ -\int_X \a^-\wedge *\a^-\cr
&=\int_X |\a^+|^2 d\m -\int_X |\a^-|^2 d\m,\cr
}}
where $*$ is the Hodge star operator and $d\m = \o^2/2!$ is the volume form.
Thus $X$ has $b_2^+ = 1$ if and only if $p_g =0$ and its intersection form is
of type $(1,b_2^-)$.

\subsec{\bf The Donaldson polynomial invariants}

Let $E$ be a complex vector bundle over $X$ with the reduction of
the structure group to $SU(2)$.
The bundle $E$ is classified by the instanton number $k$,
\eqn\aba{
k = <c_2(E), X> =\Fr{1}{8\pi^2}\int_X \tr(F\wedge F)\;\;\; \in\; \BZ^+,
}
where $\tr$ is the trace in the $2$-dimensional representation and
$\tr(\xi)^2 = -|\xi|^2$ on $\eufm{su(2)}$.
Let $\CA_k$ be the space of all connections of $E$ and
$\CA^{1,1}_k \in \CA_k$ be the  subspace consisting of the connections
whose curvatures are of type $(1,1)$, i.e.~$A\in \CA^{1,1}_k$ iff $F^{0,2}(A) =
0$.
We will call a connection $A$  holomorphic if $A\in \CA^{1,1}_k$.
Let $\CG$ be the group of the gauge transformations.
We denote $\CM_k(g)$ to the moduli space ASD connections with
respect to a (Riemann) metric $g$. We denote $\CA^*_k$
the space of irreducible connections, and thus,
$\CA^{*1,1}_k \equiv \CA^*_k \cap \CA^{1,1}_k$ and
$\CM^{*}_k(g) \equiv \CA^*_k \cap \CM_k(g)$.
The virtual complex dimension of $\CM_k(g)$  is $d = 4k -3$.

We briefly review a definition of the Donaldson polynomial invariants\foot{
There are several other more conceptually elaborated and powerful
definitions. For details, the reader can consult the excellent book \DK.
The definition of the invariants in this introduction is to emphasize
that one should also take care of the reducible ASD connections
with lower instanton numbers.}.
For  a manifold with $b_2^+ > 0$,  there are no reducible ASD connections for
a generic choice of the metric. Let $g$ be  such a  generic metric. Then
$\CM_k(g) = \CM^*_k(g)$ is a smooth manifold with actual complex dimension $d$.
Since the moduli space is rarely compact one should compactify $\CM^*_k$
to get a fundamental homology class. The Donaldson-Uhlenbeck compactification
$\overline{\CM}_k$ of $\CM^*_k$ is the closed subset of the embedding
of $\CM_k$ to the disjoint union\foot{
For manifold with $p_g=0$, it was recently shown that the  compactified moduli
space $\overline{\CM}_k$ can be identified with the total space itself \HL.},
\eqn\abd{
\bigcup_{\ell=0}^k
\CM_\ell\times \hbox{Sym}^{k-\ell}(X).
}
The compactified space includes the ASD moduli spaces with lower
instanton numbers in the lower stratas which can contain reducible
ASD connections.
To get a well-defined
invariants we should choose the metric such that it does not admit
any reducible ASD connections with instanton numbers $1,..., k$.
The genericity of the metric always means to satisfy this additional
requirement.
Let
\eqn\abe{
\m: H_2(X;\BZ) \longrightarrow H^2(\CM^*_k(g);\BZ)
}
be the Donaldson
$\mu$-map. Then we have a natural extension of $\m$
\eqn\abf{
\overline{\m}:H_2(X;\BZ) \longrightarrow H^2(\overline{\CM}_k(g);\BZ).
}
For $k > 1$ (the stable range), the compactified moduli space carries the
fundamental homology class.
The Donaldson polynomial $\overline{q}_{X,g,k}$,
\eqn\abg{
\overline{q}_{X,g,k}(\S_1,...,\S_{4k-3}) =
<\overline\m(\S_1)\smile...\smile\overline\m(\S_{4k-3}),
[\overline{\CM}_k(g)]>
}
defines the map;
$$
H_2(X,\BZ)\times\cdots\times H_2(X,\BZ) \rightarrow \BZ,
$$
which is well-defined for the  generic metric $g$.

The above definition of the Donaldson polynomial  is  basically the same as
those for manifolds with $b_2^+ > 1$. The special feature of
the manifold with $b_2^+ = 1$ is that the polynomial  actually
depends on the metric. To show that the  polynomial \abg\ is metric
independent, one should consider a smooth generic path  $g_t$ of
metrics  joining the two generic metrics and show that its value does not
change according to the variation of the metric. This amounts to show that
the one parameter family of the moduli space $\CM_k(g_t)$ or rather
its compactification $\overline\CM_k(g_t)$ does not depend on
$g_t$ at the level of homology.  This can be ensured if there is no point
in $g_t$ which admits the reducible ASD connections.

A $SU(2)$ connection is reducible if the $SU(2)$-bundle reduces
to direct sum of line bundles $E = \zeta \oplus \zeta^{-1}$. Since the bundle
$E$ is
classified by the second Chern-Class, from the Whitney formula, we have
the following condition for the bundle reduction
\eqn\abj{
k = <c_2(E), X>  = -c_1(\zeta)\cdot c_1(\zeta^{-1}) = -c_1(\zeta)\cdot
c_1(\zeta)
 = - \zeta \cdot \zeta
}
where $c_1(\zeta) = H^{2}(X,\BZ)$. Since $k$ should be positive definite
to admit ASD connection, the self-intersection of line bundle should be
negative definite to solve above equation. Let $\o_g$ be the family
of the self-dual two-forms associated with a metric  $g$.  The metric admits
reducible ASD connection if and only if
\eqn\abja{
\int_X c_1(\zeta) \wedge \o_{g} = 0.
}
If the number of self-dual harmonic form, whose
self-intersection is positive definite,  is greater
than $0$, we can avoid the reducible connection by perturbing
metric such that there are no reducible connection for generic choice of
metric. Now we consider a smooth generic path $g_t$ of the metric joining two
generic
metrics. For manifold with $b_2^+ > 1$ such a   path
can always avoid metric admitting the reducible ASD connection.
However, for manifold with $b_2^+ = 1$  there can be at least finite number of
the points in $g_t$ which admit  reducible ASD connections  since the subspace
of ASD two-forms in the space $H^2(X;\BR)$ has the codimension one.

Since we are considering the compactified moduli space,  we should also
consider  all the bundle reductions given by
\eqn\abjb{
  -k \leq \zeta\cdot \zeta \leq -1,
}
where $\zeta \in H^2(X;\BZ)$. Then, the compactified moduli space
$\overline{\CM}_k(g)$
does not contain any reducible ASD connection if and only if
$\int_X c_1(\zeta)\wedge \o_g \neq 0$ for all $\zeta$ satisfying \abjb.
A proper definition of the Donaldson polynomials requires the systematic
understanding of the appearance of the reducible ASD connection as one varies
the metric.

\subsec{\bf The chamber structure}

Let $\O_X$ be the positive cone in $H^2(X;\BR)$ defined by
\eqn\abc{
\O_X =\{\th \in H^{2}(X;\BR) | \th\cdot\th >0\}.
}
Since the intersection form $q_X$ is of type $(1,n)$ the positive cone
has two connected components. For each element $\zeta$ satisfying \abjb\
one defines the wall $W_\zeta = W_{-\zeta}$ by the intersection of the
hyperplane
$\zeta^\perp \in H^2(X;\BR)$ orthogonal to $\zeta$, i.e.,
$\zeta\cdot \zeta^\perp =0$,
with $\O_X$.
We denote $\CW_\ell$ by the collection of walls defined by
all $\zeta \in H^2(X;\BZ)$ satisfying $\zeta\cdot \zeta = -\ell$. We also
denote
the system $\overline{\CW}_k$ of walls  by
\eqn\abcd{
\overline{\CW}_k = \bigcup_{1\leq \ell \leq  k}\CW_\ell.
}
The set $\overline{\CC}^k_X$ of chambers\foot{The physicist
reader may find it easy to understand the chamber structure
by an analogy with $(1 + n)$ dimensional Minkowski space with metric
$diag(1,-1,\ldots,-1)$. This analogy is rigorous if the intersection form
is odd. The vector space $H^2(X;\BR)$ corresponds to the Minkowski
space. The intersection form is just the metric form and the positive cone
corresponds to the future light-cone. We consider the future cone which
contains
the ample cone.   An  integral class
$\zeta \in H^2(X;\BZ)$ corresponds to a vector  and its intersection number
to the norm squared of the vector in the Minkowski space. An integral class
with
negative self-intersection
number corresponds to a space-like vector.
Then the wall is defined by a space-like
hyperplane. Definitely, the hyperplane orthogonal
to a vector intersect with the time-like space if and only if the vector is
space-like.}
is the set of the connected components of $\O_X$ after removing
$\overline{\CW}_k$.
Without loss of generality one can only consider one of
the two connected components of $\O_X$ which contains the K\"{a}hler cone.
It is convenient to choose a
level set $H(q) \in \O_X$ defined by the $n$-dimensional hyperbolic space
satisfying $\theta\cdot\theta = 1$.
A metric $g$ determines a line\foot{Note that the space of self-dual harmonic
two-forms is one-dimensional.} in $\O_X$ made up of the cohomology classes
represented by  $g$-self-dual harmonic two forms $\o_g$. Let $[\o_g]$ be
the point of the line intersects with $H(q)$.  Now we can see that the
compactified
moduli space contains reducible ASD connection if and only if $[\o_g]$ lies
in one of the walls.

\ifx\answ\bigans\message{(This will come out with figures.}
\bigskip
     \epsfxsize=16.5truecm
\centerline{
\epsfbox{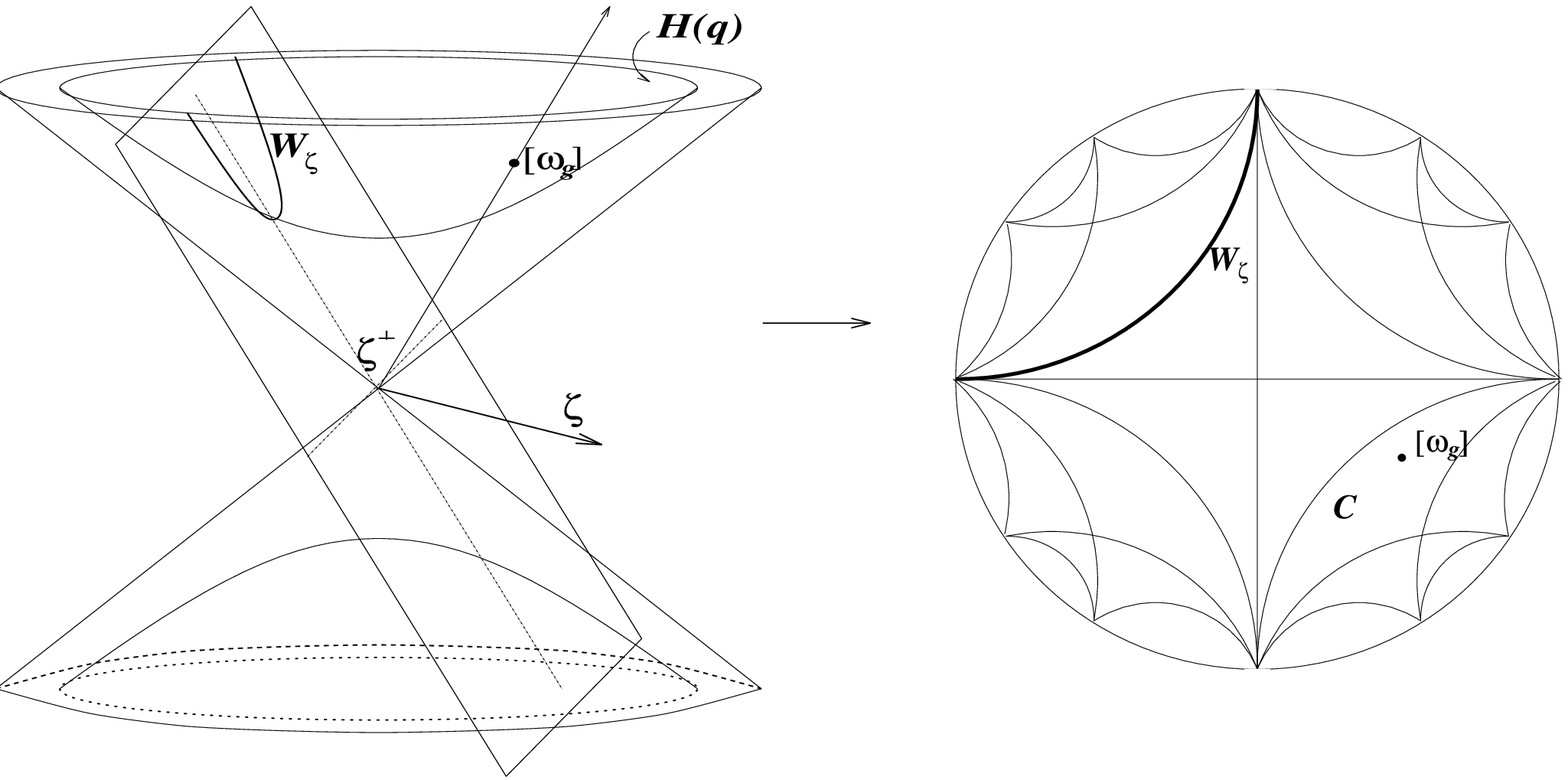}
}
\medskip
{\footskip14pt plus 1pt minus 1pt \footnotefont
{\bf Fig.~1.} A typical chamber structure \DonaldsonC\
for a manifold $X$ of type $(1,2)$ and
$k=1$ i.e.,
$\zeta\cdot \zeta = -1$. The right-hand side  is the the Poincar\'{e} model for
$H(q)$ and $C$ denote the chamber containing $\o_g$.
The pattern repeats to infinity.}
 \bigskip
\else
\fi

For a smooth path $g_t$ of metrics we have the corresponding
path $[\o_{g_t}]$ in $H(q)$.
Although the moduli space $\CM_k(g_t)$ and its compactification certainly
depends
on $t$, it may not be changed at the level of homology.
If $[\o_{g_t}]$ is contained in a chamber,  the homology class of
$ [\overline{\CM}_k(g_t)]$ does not depend on $g_t$.
On the other hand, if $\o_{g_t}$ crosses the walls, special things
happen such that the moduli space is changed even at the level of
homology. The variation of the homology class of the moduli space
is essentially  due to the appearance of the reducible ASD connection.

After picking a generic metric $g$ such that $[\o_g]$
lies in one of the chambers $C\in \overline{\CC}^k_X$ and consider
\eqn\abh{
\overline{q}_{X,g,k}(\S^{4k-3})
}
an element of $\hbox{Sym}^{4k-3}(H^2(X;\BZ))$, we can define
the map
\eqn\abi{
\overline{\Gamma}^k_X : C \rightarrow \hbox{Sym}^{4k-3}(H^2(X;\BZ)),
}
which depends only on the chamber structure $\overline{\CC}^k_X$
in $\O_X$.
The $SU(2)$-invariants of $X$ introduced by Donaldson \DonaldsonC\
and extended by Mong  \Mong\ and by Kotschick-Morgan \KotsB\KoM\
are  the  assignments
\eqn\abb{
\overline{\Gamma}^k_X :
\overline{\CC}^k_X \rightarrow \hbox{Sym}^{4k-3}(H^2(X,\BZ)),
}
The polynomial $\overline{\G}^k_X(C)$ depends only on the chamber with
following properties:

i) $\overline{\G}_X^k(-C) = -\overline{\G}_X^k(C)$

ii) If $f:X_1\rightarrow X_2$ is an orientation preserving diffeomorphism
between two
such manifolds, then $\overline{\G}_{X_1}^k(f^*(C)) =
f^*(\overline{\G}_{X_2}^k(C))$.

The computation of the Donaldson invariant amounts to determine
the invariant in a certain chamber and to find a general transition formula
its variations when $[\o_{g_t}]$ cross the walls.
The Donaldson polynomial invariants $\overline{q}_{X,g,k}(\S^{4k-3})$
may  not be well-defined if $[\o_g]$ lies one of the walls due to the
singularity in the moduli space. One can also extend the definition of the
invariants including the four-dimensional class such that
$\overline{q}_{X,g,k}(\S^{4k-3-2r}(pt)^r)$. We will denote
$\overline{\Gamma}^{k,r}_X(C)$ for the corresponding assignments which
also depend only on the chamber structure.
This extension becomes more
problematic when the moduli space has singularity.

\subsec{\bf The topological Yang-Mills theory}

To begin with, we recall the $N=2$ TYM theory on compact K\"{a}hler surfaces
\ParkA\WittenB\HPa. The theory has $N=2$ global supersymmetry whose conserved
charges
$\bs$ and $\bbs$
can be identified with the operators of $\CG$-equivariant Dolbeault cohomology
of $\CA$. The algebra for the
basic multiplet $(A^\pr,A^\ppr,\p,\bar\p,\w)$ \ParkA\ is
\eqn\aca{\eqalign{
&\bs A^\pr =-\p,\cr
&\bbs A^\pr =0,\cr
&\bs A^\ppr =0,\cr
&\bbs A^\ppr =-\bar\p,\cr
}\qquad\eqalign{
&\bs \p = 0,\cr
&\bbs\p = -i\Dp\w,\cr
&\bs\bar\p = -i\Dpp\w,\cr
&\bbs \bar\p = 0,\cr
}\qquad\eqalign{
\bbs\w=0,\cr
\bs\w=0,\cr
}}
where $A^\pr$ and $A^\ppr$ denote the holomorphic and anti-holomorphic
parts of the connection one-form $A = A^\pr + A^\ppr$,
$\p \in \O^{1,0}(\gE)$, $\bar\p \in \O^{0,1}(\gE)$ and $\w \in
\O^{0}(\gE)$. Note that $\p$ can be identified with holomorphic
(co)tangent vectors on $\CA$. The fields have additional quantum numbers
characterized by the degree $(*,*)$. The operator $\bs$ carries the degree
$(1,0)$ and $\bbs$ carries the degree $(0,1)$. Assigning the degree $(0,0)$
to the connection $A$,   $\w$ is of degree $(1,1)$.  In terms of the
equivariant
cohomology the above algebra
can be represented
as follows. We let $\O^{*,*}(\CA)$ be the Dolbeault complex on $\CA$.
Now we interpret $\hbox{Fun($Lie(\CG)$)}$ to the algebra of
polynomial functions generated by $\w^a$.
Then the desired Dolbeault model of the $\CG$-equivariant
complex is $\O^{*,*}_\CG = (\O^{*,*}(\CA)\otimes \hbox{Fun}(\CG))^\CG$.
The associated differential operators with the degrees $(1,0)$ and
$(0,1)$ are $\bs$ and $\bbs$, represented by
\eqn\acb{\eqalign{
&\bs = -\sum_i \p^i\Fr{\rd}{\rd A^{\pr i}} +i\sum_{\bar i,a}\w^a
V^{\bar i}_a \Fr{\rd}{\rd \bar\p^{\bar i}},\cr
&\bbs = -\sum_{\bar i} \bar\p^{\bar i}\Fr{\rd}{\rd A^{\ppr\bar i}}
+i\sum_{i,a}\w^a
V^{i}_a \Fr{\rd}{\rd \p^{i}},\cr
}}
where $i,\bar i$ are the local holomorphic and anti-holomorphic
indices tangent to $\CA$, respectively. We have
\eqn\acc{
\bs^2=0,\qquad \bs\bbs + \bbs\bs = -i\w^a\CL_a,\qquad \bbs^2=0,
}
Thus, $\{\bs,\bbs\} = 0$ on the $\CG$-invariant subspace
$\O^{*,*}_\CG$ of $\O^{*,*}(\CA)\otimes \hbox{Fun}(Lie(\CG))$.
{\it We define the $\CG$-equivariant Dolbeault cohomology $H^{*,*}_\CG(\CA)$
by
the pairs $(\O^{*,*}_\CG(\CA),\;\bbs)$.} It was shown that for manifold with
$p_g = 0$
the $\bs$-cohomology is isomorphic to the $\bbs$ cohomology \HPa.

The action functional of $N=2$ TYM theory can be viewed as the Dolbeault
equivariant
cohomological version of the Mathai-Quillen representative of the universal
Thom class\foot{The standard reference on the de Rham equivariant cohomology
is the book \BGV. The relation with the topological field theory was studied in
\KalkA\Hitchin\Blau. Recently, an extensive and consistent review
on the cohomological field theory  based on the equivariant de Rham cohomology
appeared \CMR. The paper \VW\ contains a self-contained introduction
on the subject as well as a generalization.
}
of the infinite dimensional bundle $\CA \rightarrow \CA/\CG$. We consider the
vector space $V$
of $\gE$-valued self-dual two-forms with a linear $\CG$ action on it. Then we
can form
a homology quotient $E_\CG = \CA \times_\CG V$.  Then there exists a
equivariant map
\eqn\acd{
F^{+}: \CA \longrightarrow V \quad\hbox{by}\quad
   A \longrightarrow F^+(A),
}
where $F^+(A)$ denotes the self-dual part of the curvature two form $F(A)$ and
defines a section $s$ of $E_\CG$.
Now, the moduli space of ASD connections is the zero set of the section $s$.
In the K\"{a}hler geometry, we can decompose the vector space $V$ into
the vector spaces $V = V^{2,0}\oplus V^{1,1}_\o\oplus V^{0,2}$ of
$(2,0)$-forms,
$(1,1)$-forms proportional to the K\"{a}hler form and $(0,2)$-forms.
Using the natural complex structure on $\CA$ induced from $X$, we can also
decompose
the section $s$ (the equivariant map) into
\eqn\ace{
\eqalign{
&F^{2,0}: \CA^\pr \longrightarrow V^{2,0}\quad\hbox{by}\quad
   A^\pr \longrightarrow F^{2,0}(A^\pr),\cr
&F^{1,1}_\o: \CA \longrightarrow V^{1,1}_\o\quad\hbox{by}\quad
   A \longrightarrow f(A^\pr,A^\ppr)\o,\cr
&F^{0,2}: \CA^\ppr \longrightarrow V^{0,2}\quad\hbox{by}\quad
   A^\ppr \longrightarrow F^{0,2}(A^\ppr),\cr
}
}
where $f(A) = \Fr{1}{2}\L F^{1,1}(A)$.

To write the Mathai-Quillen representative of the Universal Thom class, we
should introduce
the set of anti-ghost multiplets for each components of the above map.
Geometrically,
the anti-ghost multiplets are various equivariant differential forms living in
the dual vector space
of $V$ and their spectrum and algebra can be uniquely determined in terms of
the Dolbeault model of the equivariant cohomology \HPb (see also \ParkA\HPa).
This leads to   a commuting anti-ghost $\bar\w\in \O^0(\gE)$ living in the dual
vector space $V^{*1,1}_\o$  with degree $(-1,-1)$.
Then we  have  multiplet $(\bar\w,i\c^0, -i\bar\c^0, H^0)$ with
transformation laws
\eqn\acf{\eqalign{
&\bs \bar\w = -i\c^0,\cr
&\bbs \bar\w =i\bar\c^0,\cr
&\bs\bar\c^0 = H^0 -\fr{1}{2}[\w,\bar\w],\cr
&\bbs\c^0 =H^0+\fr{1}{2}[\w,\bar\w],\cr
}\qquad
\eqalign{
&\bs\c^0 = 0,\cr
&\bbs\bar\c^0 = 0,\cr
&\bs H^0 =-\fr{i}{2}[\w,\c^0],\cr
&\bbs H^0 =-\fr{i}{2}[\w,\bar\c^0].\cr
}}
We also have an anti-commuting anti-ghost $\c^{2,0}$ in the dual vector space
$V^{2,0}$  with degree $(-1,0)$ and an anti-commuting
anti-ghost $\bar\c^{0,2}$ living in the dual vector space $V^{*0,2}$ with
degree $(0,-1)$
with transformation laws
\eqn\acg{\eqalign{
&\bs\c^{2,0} = 0,\cr
&\bbs\c^{2,0} = H^{2,0},\cr
&\bs\bar\c^{0,2}=H^{0,2},\cr
&\bbs\bar\c^{0,2}=0,\cr
}\qquad
\eqalign{
&\bs H^{2,0} = -i[\w,\c^{2,0}],\cr
&\bbs H^{2,0} =0,\cr
&\bs H^{0,2} = 0,\cr
&\bbs H^{0,2} =- i[\w,\bar\c^{0,2}].\cr
}}
The action functional (the universal Thom class) is given by
\eqn\ach{\eqalign{
S =
&
-i\bs \biggl(\! \Fr{1}{h^2}\!\!\int \!\tr \bar\c^{0,2}\!\wedge \!* F^{2,0} \!
\biggr)
- i\bbs\biggl(\!\Fr{1}{h^2}\!\!\int \!\tr \c^{2,0}\!\wedge\! *
F^{0,2}\!\biggr)
- (\bs\bbs -\bbs\bs)\! \biggr(\!\Fr{1}{h^2}\!\!\int\tr \c^{2,0}\wedge\!
*\bar\c^{0,2} \!  \biggl)\cr
&
- \Fr{(\bs\bbs -\bbs\bs)}{2} \biggr( \!\Fr{1}{h^2\!}\int \tr\!\left( \bar\w f
+ \c^0\bar\c^0 \right)\o^2
\biggl).\cr
}
}
A small calculation gives
\eqn\aci{\eqalign{
S =
&\Fr{1}{h^2}\!\int_X\!\!\! \tr\biggl[
  -\Ha F^{2,0}\wedge* F^{0,2}
  +i\c^{2,0}\wedge *\Dpp\bar\p
  +i\bar\c^{0,2}\wedge *\Dp\p
  - 2i[\w,\c^{2,0}]\wedge *\bar\c^{0,2}
  \phantom{\biggr]}
  \cr
& -\biggl( \fr{1}{2}f^2
  -2i[\w,\c^0]\bar\c^0
  -\bar\c^{0} \Dp^*\p
  +\c^{0}\Dpp^*\bar\p
  -\Fr{1}{2}[\w,\bar\w]^2
 % \cr
%&
+\Ha \bar\w\left( \Da^*\Da\w -2i\L [\p,\bar\p]\right)
  \biggr)\Fr{\o^2}{2!}\biggr]\;.
  \cr
}}
where we have integrated out auxiliary fields $H^{2,0}, H^0, H^{0,2}$ and used
the K\"{a}hler identities,
\eqn\kaehler{
\Dpp^* = i[\Dp,\L], \qquad \Dp^* = -i[\Dpp,\L],
}

The bosonic kinetic terms of the action is
\eqn\acj{
S=-\Fr{1}{h^2}\int \tr \biggl(\Ha|F^+|^2 + \Ha |\Da \w|^2
%\Ha[\w,\bar\w]^2
\biggr) d\m
+ \ldots.
}
The first term is the norm square of the equivariant section
of $E_\CG$.
In the $h^2\rightarrow 0$, the dominant contribution of the path integral
comes from the
instantons $F^+(A) = 0$ and the configuration satisfying
\eqn\acl{
 \Da\w = 0. %,\quad\hbox{and}\quad    [\w,\bar\w] = 0.
}
If there is a non-zero solution $\w$ of the equation \acl\ , it means the
instanton $A$ is reducible (abelian).  Then
%Such a non-zero solution also satisfies the second
%equation since
the $SU(2)$ group reduces to the $U(1)$ subgroup,
\eqn\acm{
\w =\w_c T^3 = -\Fr{i}{2}\left(\matrix{\w_c & 0\cr 0 & -\w_c}\right) \in
\eufm{su(2)}.
}
Note also that, for reducible connection, the superpotential term
\eqn\acma{
+\Fr{1}{h^2}\int \tr \biggl(\Ha[\w,\bar\w]^2
\biggr) d\m,
}
vanishes, which corresponds to the flat direction in the physical
terms.
If the reducible instantons appear, the path integral has
additional contribution from the vector space of the
$\w$-zero-modes. Of course the moduli space of ASD connections becomes
singular. Then, the semi-classical description breaks down due to the
singularity and
the topological interpretation of the theory
can be invalidated due to the non-compactness
of the vector space of the
$\w$-zero-modes\WittenAA.
One can also view the localization of the path integral by the fixed point
locus;
\eqn\acmb{
\eqalign{
\left\{\eqalign{
\bs \psi = -i\Dp\w =0\cr
\bbs \bar\psi = -i\Dpp\w=0\cr
}\right.
\quad&\Longrightarrow\quad
\Da \w = 0
\cr
\left\{\eqalign{
&\bs\bar\chi^{0,2} =-\Fr{i}{2} F^{2,0}(A)= 0\cr
&\bbs\chi^{2,0} = - \Fr{i}{2} F^{0,2}(A)=0\cr
&\bs\bar\chi^{0} =-\Fr{i}{2} f(A) -\Ha[\w,\bar\w]= 0\cr
&\bbs\chi^{0}=-\Fr{i}{2}  f(A) +\Ha[\w,\bar\w] = 0\cr
}\right.
\quad&\Longrightarrow\quad
\left\{\eqalign{
F^+_A = 0\cr
[\w,\bar\w]^2 = 0\cr
}\right.
\cr
}}

Picking a two-dimensional class $\S \in H_2(X;\BZ)$,
one can define a topological observable
\eqn\qqa{
\mu(\S) \equiv
\Fr{1}{4\pi^2}\int_\S\tr\bigl(i\w F +\p\wedge\bar\p \bigr)
\equiv  \Fr{1}{4\pi^2}\int_X\tr\bigl(i\w F +\p\wedge\bar\p \bigr)\wedge \a_\S,
}
where $\a_\S \in H^2(X;\BZ)$ is Poincar\'{e} dual to $\S$.
This observable is the field theoretic representation of the
Donaldson's $\mu$-map \abe. One also defines the observable
$\Theta$ corresponding to the four-dimensional class, $\mu(pt)$,
\eqn\qqb{
\Theta = \Fr{1}{8\pi^2}\int_X \Fr{\o^2}{2!}\tr \phi^2.
}
Assuming that there is no reducible instanton, the correlation function
\eqn\qqc{
\left< \mu(\S_1)\cdots\mu(S_s)\Theta^r \right>
=\Fr{1}{\hbox{vol}(\CG)}\int \CD X\; e^{-S}\;\mu(\S_1)\cdots\mu(\S_s)\Theta^r,
}
 is the path integral representation
of the Donaldson invariant $\overline{q}_{k,X}(\S_1,\ldots,\S_s, (pt)^r)$.
Due to the ghost number anomaly, the correlation function \qqc\ always vanish
unless
$\hbox{dim}_\BC \CM_k = s + 2r$.

It is not entirely clear how the path integral \qqc\ takes care of the
non-compactness of the instanton moduli space.  However,
at least for manifolds with $b_2^+ \geq 3$, Witten's explicit results show that
the path integral correctly leads to the Donaldson invariants.
Clearly the path integral localizes to the moduli space of ASD connections
rather than the compactified one. We do not know any clear reasoning why the
path
integral correctly reproduce the Donaldson invariant.
One may argue that the additional space added for the compactification
does not contribute to the Donaldson invariants. In fact
more elaborated definitions of the invariants, compared to that of \abg,
clearly indicate such a property under certain conditions \DonaldsonE\DK.
For a manifold with $b_2^+=1$, on the other hand, the path integral
does not exactly represent the Donaldson invariants. This can be
easily seen if one considers the reducible connections.
Clearly the path integral localizes to the moduli space of ASD connections
rather than the compactified one. In the path integral approach, then, we only
need to worry about the reducible connections corresponding to the bundle
reduction
\abj.
In any case, one can insist that the path integral approach is well-defined
whatever properties the instanton moduli space has.  We believe
that  Witten's explicit results on the $b_2^+ \geq 3$ cases and our partial
result in this paper for $b_2^+ = 1$ support such a viewpoint.

We would like to add the following remarks.

i) The form \ach\ of the action functional has been uniquely determined.
We could correctly recover every  term in the action functional of the
$N=2$ super-Yang-Mills theory. The global supersymmetry transformation
laws and the action functional are rigorously identical to
those of the twisted theory \WittenB. On the other hand,
The usual approach based on the de Rham model of the equivariant cohomology or
the $N=1$ global supersymmetry does not leads to the complete determination
of  the anti-ghost multiplets and the correct action functional.  One should
add,
so called, projection term and non-minimal term \CMR.

ii) The usual approach to TYM theory on K\"{a}hler surface based on
the $N=1$ global supersymmetry can not explain the perturbation of
Witten, breaking the $N=2$ symmetry down to $N=1$ symmetry \WittenB\HPa.
Unfortunately, the perturbation is not applicable for manifolds with
$p_g=0$.

\subsec{\bf The holomorphic Yang-Mills theory}

An obvious way out of the difficulty of the TYM theory with the reducible
connections is to eliminate
the zero-mode of $\w$ from the  theory as originally
suggested by Witten in the two-dimensional model
of  the TYM theory \WittenC.
The remarkable fact is that his method eventually
leads to a non-abelian localization theorem of the
theory of the equivariant (de Rham) cohomology\foot{
The equivariant localization  was studied in the mathematical
literatures \DH\WittenG\ABb\KalkB\Wu\ for the abelian version and
\JK\KalkC\Verg\ for the non-abelian version. The physics oriented
reader may find the review \CMR\ the most readable. See also \BTHc}.
The HYM theory is an analogous prescription for the
Donaldson theory and it is related to a Dolbeault
equivariant cohomological version of the non-Abelian
localization theorem \ParkA\HPa.
Consequently the HYM theory is a suitable model for the
Donaldson invariants on K\"{a}hler surface
with $b^+_2 = 1$.

The basic observation is that the reduction of the path integral of the TYM
theory to the
instanton moduli space is achieved by the following two steps; i) restriction
of
$\CA$ to $\CA^{1,1}$,
ii) restriction of $\CA^{1,1}$ to the solution space of ASD connections and
reduction to the
instanton moduli space by dividing by the gauge group $\CG$.  Then the second
step can be
replaced with the symplectic reduction.  Now we  we can deform
the $(1,1)$ part of the action by the one-parameter family of the action,
\eqn\acn{
S(t) = S +  t(\bs\bbs -\bbs\bs) \biggl(-\Fr{1}{h^2}\int_M \Fr{\o^2}{2!}\, \tr
\bar\w^2\biggr),
}
where $t$ is a real positive deformation parameter.
After some Gaussian integrals we are left with\foot{Here, we choose the
delta-function gauge
for simplicity.}
\eqn\aco{\eqalign{
S(t) =
&
-i\bs \biggl(\! \Fr{1}{h^2}\!\!\int \!\tr \bar\c^{0,2}\!\wedge \!* F^{2,0} \!
\biggr)
- i\bbs\biggl(\!\Fr{1}{h^2}\!\!\int \!\tr \c^{2,0}\!\wedge\! *
F^{0,2}\!\biggr) \cr
&+ \Fr{\bs\bbs -\bs\bbs}{2}
\biggl(\Fr{1}{2h^2 t}\int_M \Fr{\o^2}{2!}\,\tr f^2\biggr).
}
}
The action functional of the HYM theory is defined by
\eqn\acp{\eqalign{
S(t)_H =
 &\Fr{1}{h^2}\!\int_X\!\!\! \tr\biggl[
  -i H^{2,0}\wedge* F^{0,2}
  -i H^{0,2}\wedge* F^{2,0}
  +i\c^{2,0}\wedge *\Dpp\bar\p
  +i\bar\c^{0,2}\wedge *\Dp\p
  \biggr]
  \cr
 &-\Fr{1}{4\pi^2}\int_X\tr\bigl(i\w F
 +\p\wedge\bar\p \bigr)\wedge\o
 -\Fr{\e}{8\pi^2}\int_X \Fr{\o^2}{2!}\,\tr\w^2
 \cr
 &+\Fr{1}{4\pi^2\e}\int_X \tr(F^{2,0}\wedge F^{0,2}
 +\Ha F^{1,1}\wedge F^{1,1})
 \;\cr &+ \Fr{\bs\bbs -\bs\bbs}{2}
 \biggl(\Fr{1}{2h^2 t}\int_M \Fr{\o^2}{2!}\,\tr f^2\biggr),
 \cr
}}
where $\e$ is positive number.

Several remarks are in order.

i) The first line of the action is identical to the part of the TYM action
(Thom class)
which leads to a clear cut reduction of $T^*\CA$ to $T^*\CA^{1,1}$ without any
quantum
correction.
Thus we can regard the theory as the one defined on $T^*\CA^{1,1}$.
Similar hybrid model of the de Rahm model was suggested in
\Thompson\ independently to \ParkB.

ii) The term
\eqn\acq{
\tilde\o\equiv \Fr{1}{4\pi^2}\int_X\tr\bigl(i\w F +\p\wedge\bar\p
\bigr)\wedge\o
}
is the equivariant extension of the K\"{a}hler form
$\Fr{1}{4\pi^2}\int_X\tr(\p\wedge\bar\p)\wedge\o$ on $\CA^{1,1}$.
It also define a two dimensional class $\mu(H)$
of the Donaldson invariants associated  to the
ample class $H$.

iii) The term
\eqn\acr{
\Theta \equiv \Fr{1}{8\pi^2}\int_X \Fr{\o^2}{2!}\,\tr\w^2
}
is the four dimensional class
of the Donaldson invariant.

iv) The term $\Fr{1}{2h^2 t}\int_M \Fr{\o^2}{2!}\,\tr f^2$ is proportional to
the norm squared $<\eufm,\eufm>$  of the moment map
$\eufm{m}:\CA^{1,1}\rightarrow \O^0({\gE})^*$,
\eqn\acs{
\eufm{m}(A) = -\Fr{1}{4\pi^2}F^{1,1}_{\!A}\wedge\o
= -\Fr{1}{4\pi^2}f\o^{2},
}
where $\O^0(\gE)^*=\O^{4}(\gE)$ denotes dual of $\O^0(\gE) = Lie(\CG)$.
Since the path integral is independent of $t$, we can set $t\rightarrow 0$, and
hence
the path integral gets contributions only from the critical set of the
function
$<\eufm{m},\eufm{m}>$,  i.e., $<f, \Da f> = 0$.

We  set $t\rightarrow \infty$ so that we can omit the equivariantly exact
form.
The resulting action,
\eqn\act{
S_H =
 -\Fr{1}{4\pi^2}\int_X\tr\bigl(i\w F^{1,1}
 +\p\wedge\bar\p \bigr)\wedge\o
 -\Fr{\e}{8\pi^2}\int_X \Fr{\o^2}{2!}\,\tr\w^2
  +\Fr{k}{\e}
 +\cdots,
 }
after the Gaussian integral over $\w$,  is identical to the physical
Yang-Mills theory restricted to the space $\CA^{1,1}$ of holomorphic
connection. It is proportional to the normed-square of moment
map up to topological terms.
The real number $\e$ corresponds to the coupling constant.
The classical equation of motion is given by
\eqn\acu{
F^{2,0}(A) = F^{0,2}(A) = 0,\qquad \Da f(A) = 0,
}
which is the Yang-Mills equation of motion on $\CA^{1,1}$.
The partition function of the HYM theory exactly reduces to
an infinite dimensional non-abelian equivariant integration formula,
\eqn\acv{\eqalign{
Z(\e,k) =e^{-\Fr{k}{\e}}\times
\Fr{1}{\hbox{vol}(\CG)}&\int_{T^*\CA^{1,1}}
\!\!\CD \! A^\pr\,\CD \! A^\ppr\,
\CD \p\,\CD \bar\p\,\CD \w\cr
&\times\exp\biggl(
\Fr{1}{4\pi^2}\int_M \tr\bigl(i\w F^{1,1} +
\p\wedge\bar\p \bigr)\wedge\o
+\Fr{\e}{8\pi^2}\int_M \Fr{\o^2}{2!}\tr\w^2\biggr).\cr
}}
{\it In the limit $\e\rightarrow 0$}, the partition function is related to
certain expectation value of the original TYM theory,
up to the exponentially small terms,
\eqn\acw{\eqalign{
Z(\e,k)
&= e^{-\Fr{k}{\e}}\biggr<\exp(\tilde\o + \e \Theta) \biggl> + \hbox{
exponentially small terms }\cr
&=e^{-\Fr{k}{\e}} \sum_{r=0}^{[(4k-3)/2]}\Fr{\e^r}{(d-2r)!r!}
\biggr<\tilde\o^{d-2r}  \Theta^r \biggl>
+ \hbox{exponentially small terms},\cr
}
}
We   assumed that the path integrals are defined with respect to
the metric whose K\"{a}hler form lies in one of the chambers.  Otherwise, as we
will show in Sect.~4, the
partition function contains a non-analytic term proportional to
$\e^{2k-3/2}$, which is the contribution due to the reducible
instantons.

There is another way of justifying that the path integral
can be expressed as the sum of contributions of
the critical points. The action functional has the
global $N=2$ supersymmetry, thus, we can use the fixed point theorem of
Witten.
The important fixed point equation is
\eqn\acx{
\bbs\w =-i\Dp\w = 0,\qquad \bs\w =-i\Dpp\w = 0.
}
This equation shows that  non-zero solutions for $\w$ (the zero-modes of $\w$)
appear for reducible connections.
By eliminating $\w$ using Gaussian integral gives
\eqn\acy{
2i f + \e \w = 0.
}
Combining the above two equations we are led to the fixed point equation $\Da f
= 0$.
{\it Thus, the zero-modes of $\w$ are no longer associated with the reducible
instanton,
rather they are mapped into higher critical points.}

We can further apply the fixed point theorem to calculate the partition
function. The path integral can be done by evaluating exactly at the fixed
point locus and by evaluating one-loop contribution of the normal modes to the
fixed point locus\WittenD. This is the basic method of our calculation.
The  partition function  of  the two-dimensional
physical Yang-Mills theory, which is the similar low dimensional cousin of HYM
theory, has been calculated also by adopting the fixed point
theorem \Ger\BTHa.

Before moving to the next section, we should add a {\it cautionary remark}
on the dual roles of $\tilde\o$ (eq.\acq).  We are interested in the variation
of $<\mu(\S)^{d-2r}(pt)^r>$
where $\S$ is an arbitrary fixed element of $H_2(X;\BZ)$
according to the changes of metric.  On the other hand, the K\"{a}hler
form $\o$ and its Poincar\'{e} dual $H$ in the action and in $\tilde\o$
vary as we change the metric.  This amounts to using the different
two dimensional classes rather than fixed one.  Thus, we should read
the relation \acw\ very carefully. We will return this in Sect.~5.

\newsec{The Path Integral}
We  use the action functional of the HYM theory in the delta function
gauge
\eqn\acp{\eqalign{
S_H =
 &\Fr{1}{h^2}\!\int_X \tr\biggl[
  -i H^{2,0}\wedge* F^{0,2}
  -i H^{0,2}\wedge* F^{2,0}
  +i\c^{2,0}\wedge *\Dpp\bar\p
  +i\bar\c^{0,2}\wedge *\Dp\p
  \biggr]
  \cr
 &-\Fr{1}{4\pi^2}\int_X\tr\bigl(i\w F^{1,1}
 +\p\wedge\bar\p \bigr)\wedge\o
 -\Fr{\e}{8\pi^2}\int_X \Fr{\o^2}{2!}\,\tr\w^2
 \cr
}}
The partition function of the HYM theory has contributions from the two
branches.

{\bf A}. The non-abelian branch : $\w_f= 0$ where the full $SU(2)$ symmetry is
restored, i.e.,
the irreducible ASD connections.

{\bf B}. The abelian branch : $\w_f \neq 0$ where the non-abelian symmetry
breaks down to abelian one,
i.e., reducible holomorphic connections including reducible ASD connections, if
any.

\noindent
Then, we can divide the partition function $Z(\e,k)$ of
$N=2$ HYM theory as the sum of contributions of the two branches,
\eqn\caa{
Z(\e,k) = Z_{{\bf A}}(\e, k) + Z_{{\bf B}}(\e, k).
}
The zero coupling limit of $Z(\e,k)$ can be identified with
the symplectic volume of the instanton moduli space $\CM_k(g)$
with respect to a K\"{a}hler metric $g$\ParkB. Of course $\CM_k(g)$ is rarely
compact.
We always assume there are no $\c^{2,0}$ and $\bar\c^{0,2}$ zero-modes.

We calculate  the partition function $Z_{{\bf B}}(\e,k)$ contributed from the
branch {\bf B} as a simple application  of Witten's  fixed point theorem
\WittenD.\foot{
On the manifold $b_2^+> 1$ the branch {\bf B} is absent generically.
This means that the deformation to HYM theory is not particulary
useful.  We do not how to evaluate the crucial branch {\bf A}.
A careful application of the abelianization techniques \JK\BTHa\BTHb\
may be used to deal with the non-Abelian branch.}

\subsec{\bf The fixed points locus}

The HYM theory has the same
global supersymmetry transformation laws as the $N=2$ TYM theory.
We only deal with the
branch {\bf B}, $\w_f \neq 0$.
In this branch the BRST fixed point  is\foot{According to our
convention of the $\tr$,
the Lie algebra generators are anti-hermitian given by
\eqn\gena{
T_a = \Fr{\s_a}{2i}, \qquad T_{\pm}=T_1 \pm i T_2,
}
with Pauli matrices $\s_a$ and
\eqn\genb{
\tr T_a T_b = -\Fr{1}{2}\d_{ab}.
}
}
\eqn\fixa{
\w_f=\w_c T_3 =constant.
}

Now we  determine the fixed point solution for the gauge connections.
They are given by the reducible (holomorphic) connections which, sometimes,
will
be  called the abelian critical points.
Consider the space $\CA_k$ of all connections of a $SU(2)$-vector-bundle
$E$ over $X$ with a given instanton number $k$. We denote $\CA^{1,1}_k$ be the
subspace which consists of holomorphic connections of $\CA_k$ and
$\CA^{*}_k$ be
the space of irreducible connections.
The space of reducible
connections is then $\CA^{1,1}_k \backslash\CA^{*1,1}_k = \CA_k
\backslash\CA^{*1,1}_k$.
A holomorphic connection $A\in \CA^{1,1}_k$ endows $E$ with a holomorphic
structure $\CE_A$. The connection $A$ is reducible if and only if
$\CE_A$ splits into the sum of holomorphic line bundles
\eqn\baa{
\CE_A = L_A\oplus L^{-1}_A
}
satisfying
\eqn\bab{
c_1(L_A)\cdot c_1(L_A)=  - k,
}
$$
H^{1,1}(X;\BZ)\times H^{1,1}(X;\BZ) \rightarrow \BZ :
$$
Since we consider the case $p_g = 0 $, the above parings become the
intersection
parings
$q_X: H^2(X;\BZ)\times H^2(X;\BZ) \rightarrow \BZ$:
\eqn\bac{
\zeta\cdot \zeta = -k,
}
where $\zeta \in H^2(X;\BZ)$.
Obviously, the solutions of the above equation
are independent of the metrics on $X$. Furthermore we always
obtain in pairs $\pm\zeta$.
{\it For the simply connected case,  that we are considering,
it is known that the reducible connection corresponding to the pairs $\pm\zeta$
is unique up to gauge equivalence class} \DK.
Thus, it is sufficient to solve the
above intersection pairings to determine the gauge equivalence class of the
abelian critical points of HYM theory.
It also follows that  the abelian fixed point locus is non-degenerate and
isolated set of points. Then, the path integral calculation based on the
fixed point theorem gives an exact answer.
We will confuse  the abelian
critical point with its associated line bundle as well as its first Chern
class.
The value of the curvature two-form
$F_{f} \in \O^{1,1}(\hbox{End}_0(\CE_{A_f})$
at the fixed point locus
\eqn\bad{
F_{f} =-2\pi i
\left(\matrix{&\zeta&0\cr
&0&-\zeta\cr}\right),
}
is determined by the first Chern class of the line bundles $\zeta$ satisfying
\bac.

The values of the other fields in the fixed point locus can be read from \aca\
and
\acg
\eqn\fixa{
\psi_f =\bar\psi_f=H^{2,0}_f = H^{0,2}_f = \c^{2,0}_f =\bar\c^{0,2}_f  = 0.
}
Of course, the values of $\chi^{2,0}$ and $\bar\chi^{0,2}$ in the fixed point
locus need not to be zero.
It is sufficient to take values in the Cartan (abelian) sub-algebra to  satisfy
$\bs H^{2,0} = -i[\w,\c^{2,0}] = 0$ or $\bbs H^{2,0} = -i[\w,\bar\c^{0,2}] =
0$.
Since we are dealing
with the manifold satisfying $H^{2,0}(X) = H^{0,2} (X) = 0$, we can set the
values zero without loss of generality.

We expand the action functional about the fixed point locus up to the
quadratic terms in the action
\eqn\badd{
S_H \approx S_f + S^{(2)},
}
where the action $S_f$ in the fixed points is given by the intersection
number
\eqn\baddc{
S_f =  \Fr{i}{2\pi}\w_c (\zeta\cdot H)  + \Fr{\e}{32\pi^2}(H\cdot H)\w_c^2,
}
for every integral cohomology classes $\zeta\in H^2(X;\BZ)$
satisfying $\zeta\cdot \zeta = - k$.

\subsec{\bf The gauge fixing}

We choose unitary gauge in which
\eqn\gauge{
\w_{\pm}=0,
}
where
\eqn\decom{
\w=(\w_c+\w_3)T_3 + \w_+T_+ + \w_-T_-.
}
In this gauge $\w$ has values on the maximal torus(Cartan sub-algebra).

By following the standard Faddev-Povov gauge fixing, we introduce a new
nilpotent BRST operator
$\d$ with the algebra
\eqn\gaugea{\eqalign{
\d \w_{\pm} = \pm c_{\pm}(\w_c + \w_3), \qquad \d c_{\pm}=0
\cr
\d \w_3 = \d \w_c =0, \qquad \d \bar{c}_{\pm}=b_{\pm},  \qquad \d b_{\pm}=0,
\cr
}}
where $c_{\pm}$ and $\bar{c}_{\pm}$ are anti-commuting ghosts and anti-ghosts
respectively, and
$b_{\pm}$ are commuting auxiliary fields.
The action for gauge fixing terms reads
\eqn\agau{\eqalign{
S_{gauge} =
  &\d\biggl[\Fr{i}{4\pi^2}\int_X i\bigl(\bar{c}_-\w_+  + \bar{c}_+\w_-
\bigr)\biggr]
 \cr
= &\Fr{1}{4\pi^2}\int_X \biggl[-\bigl(b_-\w_+  + b_+\w_-
\bigr)+\bar{c}_-(\w_c+\w_3)c_+
 -\bar{c}_+(\w_c +\w_3)c_-\biggr] \Fr{\o^2}{2!}.
 \cr
}}
The resultant action has the $\d$-BRST symmetry and, hence we can use the fixed
point theorem.
The fixed point locus for $\d$-BRST is $c_{\pm}=b_{\pm}=0$. The Gaussian
integrations over $\bar{c}$
and $c$ give
\eqn\detz{
\biggl[det_{\pm}(\w_c)\biggr]_{\Omega^0}.
}

\subsec{\bf The transverse part (I)}

The transverse part corresponding the $\pm$ part of the Lie algebra
can be easily calculated by simple Gaussian integrals.

\lin{Bosonic Sector (doublet i. e. $\pm$ part)}
The quadratic action relevant to this sector is given by
\eqn\abta{\eqalign{
S^{(2)}_{\pm}(bose) =
 &\Fr{i}{h^2}\!\int_X \biggl[
  H_+^{2,0}\wedge* \bar\rd_a A^\ppr_-
 +H_-^{2,0}\wedge* \bar\rd_a A^\ppr_+
 +H_+^{0,2}\wedge* \rd_a A^\pr_-
  +H_-^{0,2}\wedge* \rd_a A^\pr_+
  \biggr]
  \cr
 &+\Fr{i}{4\pi^2}\int_X i\w_c \bigl(A'_+\wedge A^\ppr_- {+}
 A^\ppr_+\wedge A^\pr_- \bigr)\wedge\o
  \cr
}}
where
\eqn\cova{
\rd_a \a_{\pm}=\rd \a_{\pm} \pm i A_f \a_{\pm}
}
for any doublet $\a_{\pm}$.

This is a Gaussian integral which can be evaluated by shifting the variables
\eqn\cva{\eqalign{
 A^\pr_{\pm} & \rightarrow A^\pr_{\pm} \pm \Fr{4\pi^2}{h^2 \w_c}
\rd_a^*H_{\pm}^{2,0}
 \cr
 A^\ppr_{\pm} & \rightarrow A^\ppr_{\pm} \mp \Fr{4\pi^2}{h^2 \w_c}
\bar\rd_a^*H_{\pm}^{0,2}
 \cr
}}
In terms of new variables, we have
\eqn\abtb{\eqalign{
S^{(2)}_{\pm}(bose) =
 &-\Fr{4\pi^2i}{h^4}\!\int_X \Fr{1}{\w_c}\biggl[
  \rd_a^*H_+^{2,0}\wedge* \bar\rd_a^*H_-^{0,2}+
  \bar\rd_a^*H_+^{0,2}\wedge* \rd_a^*H_-^{2,0}
  \biggr]
  \cr
 & -\Fr{i}{4\pi^2}\int_X \w_c \bigl(A^\pr_+\wedge* A^\ppr_-
   {+} A^\ppr_+\wedge* A^\pr_- \bigr),
  \cr
}}
where we have used the K\"{a}hler  identities \kaehler\ as well as
the relation \foot{This is also known as the K\"{a}hler identity.
In fact we heavily rely on many special properties of the K\"{a}hler
manifold  in the calculations, which are not valid for non-K\"{a}hler
case.}
\eqn\kid{
\int_X \alpha^{1,0}\wedge\alpha^{0,1}\wedge\o = i\int_X
\alpha^{1,0}\wedge*\alpha^{0,1},
}
and self-duality of $H$.
By the Gaussian integrals over $A$ and $H$'s, we have
\eqn\detb{
\biggl[det_{\pm}(\w_c)\biggr]^{-1/2}_{\Omega^{1,0}\oplus\Omega^{0,1}}\cdot
\biggl[det_{\pm}(\Fr{\rd_a\rd_a^*}{\w_c})\biggr]^{-1/2}_{\Omega^{2,0}}\cdot
\biggl[det_{\pm}(\Fr{\bar\rd_a\bar\rd_a^*}{\w_c})\biggr]^{-1/2}_{\Omega^{0,2}}.
}
{}

\lin{The fermionic sector (doublet i. e. $\pm$ part)}

We can compute, in a similar fashion, the contribution from the transverse
fermion doublets
\eqn\afta{\eqalign{
S^{(2)}_\pm(fermi) =
 &-\Fr{i}{h^2}\!\int_X \biggl[
   \c_+^{2,0}\wedge* \bar\rd_a\bar\p_-
  +\c_-^{2,0}\wedge* \bar\rd_a\bar\p_+
  +\bar\c_+^{0,2}\wedge* \rd_a\p_-
 +\bar\c_-^{0,2}\wedge* \rd_a\p_+
  \biggr]
  \cr
 &+\Fr{i}{4\pi^2}\int_X  \biggl[\p_+\wedge* \bar\p_- + \p_-\wedge* \bar\p_+
\biggr]
  \cr
}}
where we have used \kid\ . After changing variables by
\eqn\cvb{\eqalign{
\p_{\pm} & \rightarrow \p_{\pm} - \Fr{4\pi^2i}{h^2}\rd_a^*\c_{\pm}^{2,0}
 \cr
\bar\p_{\pm} & \rightarrow \bar\p_{\pm} -
\Fr{4\pi^2i}{h^2}\bar\rd_a^*\bar\c_{\pm}^{2,0}
 \cr
}}
we have
\eqn\aftb{\eqalign{
S^{(2)}_\pm(fermi)  =
 &-\Fr{4\pi^2i}{h^4}\!\int_X \biggl[
  \rd_a^*\c_+^{2,0}\wedge* \bar\rd_a^*\bar\c_-^{0,2}+
  \rd_a^*\c_-^{2,0}\wedge* \bar\rd_a^*\bar\c_+^{0,2}
  \biggr]
  \cr
  &+\Fr{i}{4\pi^2}\int_X  \biggl[\p_+\wedge* \bar\p_- + \p_-\wedge* \bar\p_+
\biggr]
  \cr
}}
The Gaussian integrals over $\p$ and $\c$'s give
\eqn\detc{
\biggl[det_{\pm}(\rd_a\rd_a^*)\biggr]^{1/2}_{\Omega^{2,0}}\cdot
\biggl[det_{\pm}(\bar\rd_a\bar\rd_a^*)\biggr]^{1/2}_{\Omega^{0,2}}.
}

\subsec{\bf The transverse part (II) : $U(1)$ singlets}

The remaining quadratic action  is given by
\eqn\aua{\eqalign{
S^{(t)}_3 =
 &\Fr{i}{h^2}\!\int_X \biggl[
  \Fr{1}{2}H_3^{2,0}\wedge* \bar\rd A^\ppr_3
  +\Fr{1}{2}H_3^{0,2}\wedge* \rd A^\pr_3
  -\Fr{1}{2}\c_3^{2,0}\wedge* \bar\rd \bar\p_3
  -\Fr{1}{2}\bar\c_3^{0,2}\wedge* \rd \p_3
   \biggr]
  \cr
 &+\Fr{1}{4\pi^2}\int_X \biggl[\Fr{i}{2}\w_3 \bigl(\rd  A^\ppr_3 + \bar\rd
A^\pr_3 \bigr)
 +\p_3\bar\p_3 \biggr]\wedge\o
  +\Fr{\e}{8\pi^2}\int_X \Fr{\o^2}{2!}\,\Fr{1}{2}\w_3^2.
 \cr
}}
We will show that the bosonic and fermionic contributions in this part cancel
exactly each other and so give trivial result.
Note that every field in here doesn't contain any zero mode for $\rd$ and
$\bar\rd$ operator.
We explicitly decomposed $T_3$ component of $\w$ as zero mode $\w_c$ and
non-zero mode $\w_3$ from the beginning and similarly for the $U(1)$
connections
as the topologically non-trivial part, which is contained in  the action
$S_f$,
and trivial one. Furthermore
as $h^{1,0}=h^{0,1}=h^{2,0}=h^{0,2}=0$,
where $h^{p,q}$ denotes the Hodge number, in our considerations all other
fields
do not have any zero-mode\foot{Actually we can remove the condition
$h^{1,0}=h^{0,1}=0$.  We leave this as an exercise.}.

{}From $H_3^{2,0}$ and $H_3^{0,2}$ integrations, we get the delta function
constraints
\eqn\eoma{
  \bar\rd A^\ppr_3 = 0, \qquad  \rd A^\pr_3 =0.
}
Similarly the  $\c^{2,0}_3$ and $\bar\c^{0,2}_3$ integrations give the
constraints
\eqn\eomc{
 \bar\rd \bar\p_3=0 , \qquad       \rd \p_3=0.
}
{}From the K\"{a}hler identities \kaehler, we have
\eqn\eoma{
\eqalign{
\rd A^\pr_3 = \rd^* A^\pr_3 =0,\cr
\bar\rd A^\ppr_3 = \bar\rd^* A^\ppr_3 =0,\cr
}\qquad
 \eqalign{
\rd \psi_3 = \rd^* \psi_3 =0,\cr
\bar\rd\psi_3 = \bar\rd^* \bar\psi_3 =0.\cr
}}
This shows that all of them are harmonic one-forms which implies
they actually vanishes.  Thus the delta function constraints \eoma\ and \eomc\
are equivalent to the delta function constraint,
$$
\prod_{x \in
X}\delta(A^\pr_3(x))\delta(A^\ppr_3(x))\delta(\psi_3(x))\delta(\bar\psi_3(x)).
$$
Thus the integrations over $A^\pr_3,A^\ppr_3,\psi_3,\bar\psi_3$
just lead to an unity.
Finally, the integration of non-constant mode $\w_3$ is  trivial
as only the quadratic term left in the action.

By combining \detz\ , \detb\ and \detc\ , we have the contributions of the
transverse degrees of
freedom as follows:
\eqn\detd{\eqalign{
&\biggl[det_{\pm}(\w_c)\biggr]_{\Omega^0}\cdot
 \biggl[det_{\pm}(\w_c)\biggr]^{-1/2}_{\Omega^{1,0}\oplus\Omega^{0,1}}\cdot
 \biggl[det_{\pm}(\w_c)\biggr]^{-1/2}_{\Omega^{2,0}\oplus\Omega^{0,2}}
\cr
=&\biggl[det_{\pm}(\w_c)\biggr]^{1/2}_{(\Omega^0\ominus\Omega^{1,0}\oplus\Omega^
{2,0})
 \oplus(\O^0\ominus\Omega^{0,1}\oplus\Omega^{0,2})}
 \cr
=&\biggl[det(\w_c)\biggr]^{index\bar\rd_{(+)} + index \bar\rd_{(-)}} =
\w_c^{2(1-h^{0,1} + h^{0,2})-4k}
 =\w_c^{2-4k} ,
\cr
}}
where we used the index theorem of the twisted Dolbeault operators.

\newsec{\bf The Partition Function}

The path integral of the previous section shows that
 partition function essentially reduces
 to the following expression
\eqn\nna{
Z_{\bf B}(\e,k) = \sum_{ \zeta_i\cdot\zeta_i = -k}
\int^{\infty}_{-\infty} d\w \Fr{1}{\w^{4k-2}}
\exp\biggl[-\Fr{i}{2\pi}\w(\zeta_i\cdot H)
-\Fr{\e}{32\pi^2} \w^2 (H\cdot H)\biggr],
}
where we omitted the superscript $c$ from $\w$.
In this section we examine the small coupling behavior of the above partition
function \foot{The similar asymtotic estimation of the finite dimensional
integral \nna\ was discussed by Wu \Wu\ in his study of the abelian
localization as a special case of the non-abelian localization.
Note that the branch {\bf B} is governed by abelian localization
rather than the sophiscated non-abelian localization.
}
and its variations according to the metric.

\subsec{\bf The Small Coupling Behavior}

The above equation has pole at $\w =0$, which can be removed by deforming the
contour from $\BR$ to $P_\pm = \{Im\,\w = \pm \kappa\}$ and taking the limit
$\kappa \rightarrow 0$. Of course the integral should be independent to the
contours. For
$\w \in P_\pm$, we can write
\eqn\nnb{\eqalign{
Z_{\bf B}&(\e,k)\cr
&= -\sum_{\zeta_i\cdot \zeta_i=-k}
\!\!\!\Fr{1}{(4k-3)!}\int_{0}^{\infty}\!\! ds\, s^{4k-3}\int_{P_\pm}\!
d\w \exp\biggl[\pm i\w s -\Fr{i}{2\pi}\w(\zeta_i\cdot H)
-\Fr{\e}{32\pi^2}\w^2(H\cdot
H)\biggl]\cr
&=-\sum_{ \zeta_i\cdot \zeta_i = -k}
\!\!\!\Fr{1}{(4k-3)!}\sqrt{\Fr{32\pi^3}{\e(H\cdot H)}}
\int_0^\infty \!\!ds\, s^{4k-3}\exp\biggr[-\Fr{8\pi^2}{\e(H\cdot H)}\left(
s\pm \Fr{\zeta_i\cdot H}{2\pi}\right)^2\biggr].\cr
}}

We choose the $+$ sign and decompose the above equation as
\eqn\nnc{\eqalign{
Z_{\bf B}(\e,k)
=
&-\sum_{\zeta_i\cdot H < 0}
\Fr{1}{(4k-3)!}\sqrt{\Fr{2\pi}{\e^\pr}}\int_{0}^{\infty} ds\, s^{4k-3}
e^{-\Fr{1}{2\e^\pr}\left(s + \Fr{\zeta_i\cdot H}{2\pi}\right)^2}\cr
&-\sum_{\zeta_i\cdot H = 0}
\Fr{1}{(4k-3)!}\sqrt{\Fr{2\pi}{\e^\pr}}\int_{0}^{\infty} ds\, s^{4k-3}
e^{-\Fr{1}{2\e^\pr}s^2}\cr
&-\sum_{ \zeta_j\cdot H > 0}
\Fr{1}{(4k-3)!}\sqrt{\Fr{2\pi}{\e^\pr}}\int_{0}^{\infty}ds\,
s^{4k-3} e^{-\Fr{1}{2\e^\pr}\left(s + \Fr{\zeta_j\cdot H}{2\pi}\right)^2},\cr
}}
where we set $\e^\pr = \e(H\cdot H)/16\pi^2$.
One can easily see the last term vanishes in the limit $\e^\pr \rightarrow 0$
up
to exponentially small
term as follows;
\eqn\nnp{\eqalign{
\sum_{\zeta_i\cdot H > 0}&
\Fr{1}{(4k-3)!}\sqrt{\Fr{2\pi}{\e^\pr}}\int_{0}^{\infty} ds\, s^{4k-3}
e^{-\Fr{1}{2\e^\pr}\left(s + \Fr{\zeta_i\cdot H}{2\pi}\right)^2} \cr
\leq&\sum_{ \zeta_j\cdot H > 0}
\Fr{1}{(4k-3)!}\sqrt{\Fr{2\pi}{\e^\pr}}\int_{0}^{\infty}ds\,
s^{4k-3} e^{-\Fr{1}{2\e^\pr}\left(s^2 +\left(\Fr{\zeta_j\cdot
H}{2\pi}\right)^2\right)} \cr
=&\sum_{ \zeta_j\cdot H > 0}
\Fr{1}{(4k-3)!}\sqrt{\Fr{2\pi}{\e^\pr}}\int_{0}^{\infty}ds\,
s^{4k-3} e^{-\Fr{1}{2\e^\pr}s^2} \cdot e^{-\Fr{1}{\e}\cdot
\Fr{2(\zeta_j\cdot H)^2}{H\cdot H}} \cr
=&O\left(e^{-\Fr{\d^2}{\e}}\right), \cr
}}
where $\d^2=Min_j(\Fr{2(\zeta_j\cdot H)^2}{H\cdot H})$.
Similarly we can extract the polynomial dependency on $\e$, apart from the
exponentially small
terms, from the first term;
\eqn\nnq{\eqalign{
-\sum_{\zeta_i\cdot H < 0}&
\Fr{1}{(4k-3)!}\sqrt{\Fr{2\pi}{\e^\pr}}\int_{0}^{\infty} ds\, s^{4k-3}
e^{-\Fr{1}{2\e^\pr}\left(s +\Fr{\zeta_i\cdot H}{2\pi}\right)^2}\cr
=&-\sum_{ \zeta_j\cdot H < 0}
\Fr{1}{(4k-3)!}\sqrt{\Fr{2\pi}{\e^\pr}}\int_{-\infty}^{\infty}ds\,
s^{4k-3} e^{-\Fr{1}{2\e^\pr}\left(s +\Fr{\zeta_j\cdot H}{2\pi}\right)^2}
 + O\left(e^{-\Fr{\d^2}{\e}}\right)\cr
=&-\sum_{ \zeta_j\cdot H < 0}
\Fr{1}{(4k-3)!}\sqrt{\Fr{2\pi}{\e^\pr}}\int_{-\infty}^{\infty}ds^\pr\,
\left(s^\pr -\Fr{\zeta_j\cdot H}{2\pi}\right)^{4k-3} e^{-\Fr{1}{2\e^\pr}s^{\pr
2}}
 + O\left(e^{-\Fr{\d^2}{\e}}\right)\cr
=& \sum_{\zeta_i\cdot H < 0}
\sum_{r =0}^{[d/2]}
\Fr{{2}\pi }{(d-2r)! \,r!\, 2^r }\left(\Fr{\zeta_i\cdot H}{2\pi}\right)^{d-2r}
(\e^\pr)^r
+ O\left(e^{-\Fr{\d^2}{\e}}\right)\cr
=&\Fr{1}{(2\pi)^{d-1}}\sum_{\zeta_i\cdot H < 0}\sum_{r=0}^{[d/2]}
\Fr{\e^r}{(d-2r)!\,r!\, 2^{3r} }\left(\zeta_i\cdot H\right)^{d-2r}
\left({H\cdot H}\right)^r
+ O\left(e^{-\Fr{\d^2}{\e}}\right),\cr
}}
where $d=4k-3$.  Assume that there are no
reducible ASD connections,
i.e., no line bundle $\zeta_i$ with $\zeta_i\cdot H = 0$, such that
the second term absent.
Then,
our result
for $Z^\pr(\e,k )$ is
\eqn\cab{
Z_{\bf B}(\e, k) =  \Fr{1}{(2\pi)^{d-1}}
\sum_{\zeta_i\cdot H < 0}
\sum_{r =0}^{[d/2]}
\Fr{\e^r}{(d-2r)!\,r!\, 2^{3r}}\left(\zeta_i\cdot H\right)^{4k-3-2r}
\left({H\cdot H}\right)^r
+ O\left(e^{-\Fr{\d^2}{\e}}\right),
}
where the summation runs over every divisor satisfying
$\zeta_i \cdot \zeta_i = -k$ and $\zeta_i \cdot H < 0$.
If we choose the $-$ sign in the beginning we have the same pattern of the
asymptotic
behaviors and the divisors satisfying $\zeta_i \cdot H > 0$ only contributes.
Since the solutions of $\zeta_i\cdot\zeta_i = -k$
always arise in pairs $\pm \zeta_i$, we have the same
result.

\subsec{\bf The Variation of the Partition Function}

We have seen that the Abelian critical points of HYM theory
are the  reducible  connections which can be uniquely
determined by the $2$-dimensional integral classes (or line bundles)
$\zeta_i$ satisfying $\zeta_i\cdot \zeta_i = -k$. Clearly, the notion of the
reducible connections is metric independent. On the other hand,
the notion of the reducible ASD connections is evidently metric dependent.  An
Abelian critical point $\zeta$ is ASD if and only if $\zeta\cdot H = 0$
where $H$ is Poincar\'{e} dual to the K\"{a}hler form associated with the given
metric. Equivalently, a reducible connection $A$ is ASD if and only if
the degree of the associated holomorphic line bundle $\zeta$ is zero,
 \eqn\bae{
\hbox{deg}(\zeta) = \int_X  c_1(\zeta)\wedge\o = \zeta\cdot H
=\Fr{i}{2\pi}\int_X F_c\wedge \o
=\Fr{if_c(A)}{2\pi} \int_X \o \wedge\o.
}
The degree depends only on the cohomology classes of the first Chern class
and of the K\"{a}hler form.
The value $f_c(A_f)$ of the critical points $\Da f_c(A) = 0$ is also determined
by the degree and the self-intersection number of the K\"{a}hler form.
Note that the critical points of the moment map $\eufm{m}(A)$ are also
determined by the same data.
We call a reducible connection as a higher critical point
if its degree is non-zero.

Before going on, we should remark that the structure of the abelian critical
points is isomorphic to the chamber
structure determined by the system $\CW_k\subset \overline{\CW_k}$ of walls
in the positive cone.
In particular, for each line bundle associated with an abelian
critical point $\zeta$, there is an associated wall $W_\zeta$ in the positive
cone.
Let $H_+$ and $H_-$ be ample divisor lying in the
two chambers $C_+$ and $C_-$ separated by the single wall $W_\zeta$. We assume
that
 $\zeta$ has negative degree with respect
to ample divisor $H_+$ lying in the chamber $C_+$, i.e., $ \zeta \cdot H_+ <
0$.
Let $H_0$ denote ample divisor lying on the wall. If we change
the metric, such that the ample divisor $H$ starts from $C_+$, crosses the wall
and goes to the chamber $C_-$,
we have\foot{The walls can intersect with each other, we assume that our
path of metric does not cross the intersection region.}
\eqn\baf{
 \zeta \cdot H_+ < 0 ,\qquad \zeta \cdot H_0 = 0 ,\qquad \zeta \cdot H_- > 0.
}
Now it is clear what we have actually done by deforming TYM theory to HYM
theory. We mapped all the possible reducible instantons, which appear as we
vary
the K\"{a}hler metric in every possible way, to the abelian critical points
of the HYM theory.

In the limit $\e\rightarrow 0$, the partition function of HYM theory is
such that we sum up only critical points
with negative degree. As one can easily see, we lost a critical point $\zeta$
with negative degree by passing through the wall $W_\zeta$.
On the other hand,  there is also another abelian  critical point $-\zeta$
which defines the
same wall $W_\zeta$ with $\zeta$. Thus,   we get a new critical point
$-\zeta$ of negative degree instead.
 We will collectively denote
the ample classes lying in a $C_\pm$ by $H_\pm$.
We change the metric such that our ample class $H$ crosses just {\it one wall}
$W_\zeta$, defined by a certain divisor $\zeta$  satisfying $\zeta \cdot H_+ <
0$
and $\zeta\cdot\zeta = -k$,
and moves to another chamber $C_-$.
Then we have $\zeta\cdot H_- > 0$ by definition.
This also amounts that no other line
bundles change the sign of their degrees.
The contribution of $\zeta$ to the partition function
$Z_{\bf B}(\e,k)(C_+)$
is given by
\eqn\ccaf{\Fr{{1}}{(2\pi)^{4k-4}}
\sum_{r =0}^{[d/2]}
\Fr{\e^r}{(d-2r)!\,r!\,2^{3r} }\left(\zeta\cdot H_+\right)^{4k-3-2r}
\left({H_+\cdot H_+}\right)^r
+ O\left(e^{-\Fr{\d^2}{\e}}\right),
}
while  $Z_{\bf B}(\e,k)(C^-)$ no longer receives contributions from $\zeta$.
On the other hand,
$Z_{\bf B}(\e,k)(C_-)$
receives contributions from  $-\zeta$, since $-\zeta\cdot H_- < 0$, given by
\eqn\ccag{
\Fr{{1}}{(2\pi)^{4k-4}}
\sum_{r =0}^{[d/2]}
\Fr{\e^r}{(d-2r)!\,r!\, 2^{3r}}\left(-\zeta\cdot H_-\right)^{4k-3-2r}
\left({H_-\cdot H_-}\right)^r
+ O\left(e^{-\Fr{\d^2}{\e}}\right),
}
Without loss of generality, we will use a normalization  $H\cdot H =1$,
i.e., $H_+\cdot H_+=H_-\cdot H_-=1$.

\medskip\subsec{\bf The non-analytic part}

Up to now, we have considered the case that the metric does not admit reducible
ASD
connections. Now we allow the K\"{a}hler metric to lie on one of the walls
$\CW_k$ such that there is a reducible ASD connection.
We define the multiplicity $m$ of the reducible
instanton by the number of the walls intersect at the point where the
K\"{a}hler metric lies on.
Then the partition function \caa\ contains a non-analytic term
which is not a polynomial of $\e$.
Now the partition function \cab\ has additional contribution
given by
\eqn\nnr{\eqalign{
-\sum_{\zeta_i\cdot H = 0}&
\Fr{1}{(4k-3)!}\sqrt{\Fr{2\pi}{\e^\pr}}\int_{0}^{\infty} ds\, s^{4k-3}
e^{-\Fr{1}{2\e^\pr}s^2}\cr
=&-
\Fr{m}{(4k-3)!}\sqrt{\Fr{2\pi}{\e^\pr}}\int_{0}^{\infty} ds\, s^{4k-3}
e^{-\Fr{1}{2\e^\pr}s^2}\cr
=&-\Fr{(2\pi)^{1/2}m}{(4k-3)!!}(\e^\pr)^{2k-3/2}, \cr
}}
leading to the
 modified partition function
\eqn\caba{\eqalign{
Z^\pr(\e, k) = & \Fr{{1}}{(2\pi)^{4k-4}}
\sum_{\zeta_i\cdot H < 0}
\sum_{r =0}^{[d/2]}
\Fr{\e^r}{(d-2r)!\,r!\, 2^{3r} }\left(\zeta_i\cdot H\right)^{4k-3-2r}
 \cr
&-\Fr{1}{(2\pi)^{4k-4}}\Fr{m }{(2\pi)^{1/2}(4k-3)!!} (\e/2)^{2k-3/2}\cr
&+ O\left(e^{-\Fr{\d^2}{\e}}\right).\cr
}
}
Our calculation precisely shows
that the non-analytic term is the contribution of the reducible instanton.
The similar phenomenon has been observed in the two-dimensional version
of the TYM theory by Witten.\foot{We refer the readers to the remarks
in Sect.~4.4 of ref.~\WittenC.}
%The non-analytic behavior of the partition function coincides
%with Donaldson's observation that his invariants
%$\overline{q}_{k,X}(\overbrace{\S,..,\S}^{d-2r},\overbrace{pt,..,pt}^r)$
%are well defined polynomials only for small enough $r$.
%Anyway the field theoretic method is well defined in any case.

\newsec{The Relations with the TYM Theory}

Now we are going to extract the expectation values of topological observables
in the TYM theory. To obtain the precise relations with the TYM theory when we
vary
the metric, it is conceptually more clear and convenient to
use the expectation values of the topological observables rather than the
partition
function itself. The HYM theory has the same set of topological observables
\qqa\qqb\ with the TYM theory. The relation \acw\
between  certain correlation function of TYM and the partition of HYM theory
can be generalized \WittenC\ParkB.
Picking two-dimensional classes $\S_i \in H_2(X;\BZ)$,
the expectation value $\left<\prod_{i=1}^d \mu(\S_i)\right>^\pr$ evaluated
in the HYM in the limit $\e\rightarrow 0$ is related to that of TYM theory by
the formula
\eqn\pqb{
\left<\prod_{i=1}^d \mu(\S_i)\right>^\pr
=  \left<\prod_{i=1}^d \mu(\S_i)\right> +  O\left(e^{-\Fr{\d^2}{\e}}\right).
}
This can be further generalized to include the four-dimensional
class $\Theta$
\eqn\pqc{
\left<\prod_{i=1}^{d-2r} \mu(\S_i)\right>^\pr
= \sum_{2m +n =2r} \Fr{\e^m}{m!n!}\left<\prod_{i=1}^{d-2r}
\mu(\S_i)\,\tilde\o^n\Theta^m\right>
          + O\left(e^{-\Fr{\d^2}{\e}}\right).
}
Thus it is sufficient to calculate \pqc\  to determine the general
expectation values of TYM theory for a simply connected
manifold.\foot{Actually,
it is sufficient to determine the term of power $\e^r$ in the RHS of \pqc.}

The expectation value of the observables in HYM theory
can be also written as the sum of the contributions
of the two branches {\bf A} and {\bf B};
\eqn\qqc{
\left<\prod_{i=1}^{d-2r} \mu(\S_i)\right>^\pr
= \left<\prod_{i=1}^{d-2r} \mu(\S_i)\right>^\pr_{\bf A} +
\left<\prod_{i=1}^{d-2r} \mu(\S_i)\right>^\pr_{\bf B}.
}
We should emphasize here that, in the HYM theory, all the terms that do not
vanish exponentially must
be interpreted as the contributions of the ASD connections.\foot{
This statement is originated from \WittenC.}
That is, $<\cdots>^\pr_{\bf B}$ for the branch ${\bf B}$
is also the contributions of ASD connections up to the exponentially
small terms for $\e\rightarrow 0$.  Clearly,  $<\cdots>^\pr_{\bf A}$
for the branch ${\bf A}$ does not contain any exponentially
small term.  Consequently, we can divided any expectation value
of TYM theory as
\eqn\qzz{
\bigl<\prod_{i=1}^{d-2r} \mu(\S_i)\,\Theta^r\bigr>
=\bigl<\prod_{i=1}^{d-2r} \mu(\S_i)\,\Theta^r\bigr>_{{\bf A}}
+\bigl<\prod_{i=1}^{d-2r} \mu(\S_i)\,\Theta^r\bigr>_{{\bf B}}
}
where
\eqn\qzza{\eqalign{
\bigl<\prod_{i=1}^{d-2r} \mu(\S_i)\,\Theta^r\bigr>^\pr_{{\bf A}}
&=\bigl<\prod_{i=1}^{d-2r} \mu(\S_i)\,\Theta^r\bigr>_{{\bf A}},
\cr
\bigl<\prod_{i=1}^{d-2r} \mu(\S_i)\,\Theta^r\bigr>^\pr_{{\bf B}}
&=\bigl<\prod_{i=1}^{d-2r} \mu(\S_i)\,\Theta^r\bigr>_{{\bf B}}
+ O\left(e^{-\Fr{\d^2}{\e}}\right). \cr
}}
If we consider manifold with $b_2^+ \geq 3$,
the contribution of the branch {\bf B} is generically absent for both
HYM and TYM theories.

In Sect.~5.1, the contribution of the branch ${\bf B}$ to the expectation
values
is evaluated using the similar method  as for the partition function
$Z_{\bf B}(\e,k)$. However, we do not know how to calculate the
contribution of the branch ${\bf A}$.  It is quite natural to believe that
the Seiberg-Witten monopole invariants correspond to the branch
${\bf A}$.  In Sect.~5.2, we briefly review some known properties
of the Seiberg-Witten invariants for manifolds with $b_2^+ =1$.
Then, we study the variation of the expectation values of TYM
theory in Sect.~5.3.

\medskip\subsec{\bf The  branch ${\bf B}$}

As usual, we choose a generic metric $g$ which does
not admit the zero-modes of  $\bar\c^{0,2}$.
That is, the second instanton cohomology group is
trivial. Then the only source of the singularities
in the moduli space of ASD connection $\CM_k(g)$ is
the reducible instantons

The value $\left<\prod_{l=1}^{d-2r} \mu(\S_l)\right>^\pr_{\bf B}$
for $r = 0,1,\ldots, [d/2]$
can be evaluated using the fixed
point theorem as the partition function $Z_{\bf B}(\e,k)$ is. The result
can be readily obtained by the effective partition function \nna\ as
\eqn\qqh{
\left<\prod_{l=1}^{d-2r} \mu(\S_l)\right>^\pr_{\bf B}
= \sum_{ \zeta_i\cdot\zeta_i = -k}
\prod_{l=1}^{d-2r} \left(-\Fr{i(\zeta_i\cdot \S_l)}{2\pi}\right)
\!\int^{\infty}_{-\infty} \!\!d\w \Fr{1}{\w^{2r+1}}
\exp\biggl[-\Fr{i}{2\pi}\w(\zeta_i\cdot H) -\Fr{\e}{32\pi^2} \w^2 \biggr]
}
We can study the $\e\rightarrow 0$ limit using the
methods in Sect.~4.1.,
\eqn\qqha{\eqalign{
\left<\prod_{l=1}^{d-2r} \mu(\S_l)\right>^\pr_{\bf B}
=&\sum_{\zeta_i\cdot H < 0}
\prod_{l=1}^{d-2r} \left(\Fr{\zeta_i\cdot \S_l}{2\pi}\right)
\Fr{1}{(2r)!}\sqrt{\Fr{2\pi}{\e^\pr}}\int_{0}^{\infty} ds\, s^{2r}
e^{-\Fr{1}{2\e^\pr}\left(s + \Fr{\zeta_i\cdot H}{2\pi}\right)^2}
\cr
&+\sum_{\zeta_i\cdot H = 0}
\prod_{l=1}^{d-2r} \left(\Fr{\zeta_i\cdot \S_l}{2\pi}\right)
\Fr{1}{(2r)!}\sqrt{\Fr{2\pi}{\e^\pr}}\int_{0}^{\infty} ds\, s^{2r}
e^{-\Fr{1}{2\e^\pr}s^2}
\cr
&+\sum_{ \zeta_j\cdot H > 0}\prod_{l=1}^{d-2r}
\left(\Fr{\zeta_i\cdot \S_l}{2\pi}\right)
\Fr{1}{(2r)!}\sqrt{\Fr{2\pi}{\e^\pr}}\int_{0}^{\infty}ds\,s^{2r}
e^{-\Fr{1}{2\e^\pr}\left(s + \Fr{\zeta_j\cdot H}{2\pi}\right)^2},\cr
}}
The third term above vanishes exponentially fast for  $\e\rightarrow 0$.

It is amusing to note that the second term in \qqha, corresponding
to the contributions of the reducible ASD connections, always vanish.
If we have a solution $\zeta$ for $\zeta\cdot\zeta = -k$ and $\zeta\cdot H=0$,
we always have another solution $-\zeta$. Since $d-2r =4k-3-2r$
is always an odd integer, their contributions cancel with each other.
Thus, $\left<\prod_{l=1}^{d-2r} \mu(\S_l)\right>^\pr_{\bf B}$ is well-defined
even in the presence of the reducible ASD connections.

The first term can be calculated in the same way as \nnq, which gives
\eqn\qqd{
\left<\prod_{l=1}^{d-2r} \mu(\S_l)\right>^\pr_{\bf B} =
\Fr{1}{(2\pi)^{d-1}}\sum_{\zeta_i\cdot H < 0}
\Fr{\e^r}{2^{3r}\, r!}\prod_{l=1}^{d-2r}(\zeta_i\cdot \S_l)
+ \hbox{lower orders in $\e$}
+ O\left(e^{-\Fr{\d^2}{\e}}\right).
}
 From the relation \pqc, \qzz\ and \qzza, we have
\eqn\qqd{
\left<\prod_{l=1}^{d-2r} \mu(\S_l)\,\Theta^r\right>_{\bf B}
= \Fr{1}{(2\pi)^{d-1}}
\sum_{\matrix{{}_{\zeta_i\cdot H < 0,} \cr  {}_{\zeta_i\cdot\zeta_i = -k}\cr}}
 \Fr{1}{ 2^{3r}}   \prod_{l=1}^{d-2r}(\zeta_i\cdot \S_l).
}
Then, the general form of  the expectation values of the topological
observables
of TYM theory is given by
\eqn\qqe{
\left<\prod_{l=1}^{d-2r} \mu(\S_l)\,\Theta^r\right>
= \left<\prod_{l=1}^{d-2r} \mu(\S_l)\,\Theta^r\right>_{\bf A}
+ \Fr{1}{(2\pi)^{d-1}}
\sum_{\matrix{{}_{\zeta_i\cdot H < 0,} \cr  {}_{\zeta_i\cdot\zeta_i = -k}\cr}}
 \Fr{1}{2^{3r}}   \prod_{l=1}^{d-2r}(\zeta_i\cdot \S_l).
}

\medskip\subsec{\bf The Seiberg-Witten invariants}

One of the open problem in our approach is to uncover
the genuine diffeomorphism invariants. This is because we computed
the path integral only for
one of the two branches of the fixed points. Clearly, the branch {\bf B},
which we have calculated, is generically absent for manifold with $p_g > 1$.
For the fixed point $\w = 0$ (the branch {\bf A}), we have the full
$SU(2)$ symmetry and the path integral has contribution
from the irreducible instantons.  There is no reason that the contribution
would
vanish. Unfortunately, the path integral for this branch reduces to
a formal expression such as one integral over the moduli space of irreducible
instantons. Thus, we need an alternative approach to deal with this
branch.

Note that the limit $t \rightarrow 0$ in eq.(2.29) can also be viewed as
the limit $h^2\rightarrow \infty$ with $t$ fixed. In this limit, the
semi-classical analysis is invalidated. At this point we can utilize the
fundamental results of Seiberg-Witten on the strong coupling behaviour of the
untwisted $N=2$ super-Yang-Mills theory \SWa\SWb\WittenE. Their result can be
essentially summarized by  the quantum moduli space parametrized
by a complex variable $u$, which corresponds to the observable $\Theta$
in the twisted theory. Classically, there is a singularity at the origin
$u=0$ where the full $SU(2)$ symmetry is restored.
Quantum mechanically, the complex $u$-plane has two singularities
at $u =\pm 1$. The singularities in that plane
represent the appearance of new massless particles. For manifolds
with $p_g > 1$ one
can introduce a perturbation  utilizing holomorphic two-forms
such that the only contribution comes from the two singular points.
Such an effective low energy theory turns out to be an $N=2$ super-Maxwell
theory coupled with hyper-multiplet. One can twist this theory
and the resulting theory gives the dual description of the
Donaldson invariants\WittenE.\foot{The Seiberg-Witten monopole equation
is the close cousin of the equation appeared in the Vafa and Witten's paper
on a twisted $N=4$ super-Yang-Mills theory \VW. The simlarity has an obvious
origin since the $N=4$ theory can be viewed as an $N=2$ theory coupled
with a $N=2$ matter multiplet in the adjoint represenation. The similar
equation also appears in \FA, although we do not know the origin of
the similarity.  We would like to mention that  the general  $N=2$
super-Yang-Mills
theory with hypermultiplets can be twisted to define a set of topological field
theories which lead to certain non-abelian version of the Seiberg-Witten
monopole equation \HPP.}

For manifold with $p_g=0$, no such perturbation  is
possible and one should integrate over the whole complex plane \WittenE.\foot{
It was noticed that the Donaldson invariants and Seiberg-Witten
invariants are very different for manifold with $b_2^+ =1$.
There is no mystery in it  since Seiberg-Witten
monopole invariant is one of two parts of the Donaldson invariant.
}
The extra contributions
from the generic points of the quantum moduli space are the
contributions of the abelian instantons. What we have calculated in this
paper is essentially such contributions.
We can view the original Seiberg-Witten monopole invariants
as the contribution of the branch ${\bf A}$.

The Seiberg-Witten invariants can be viewed as the pairs
$(x, n_x)$ where $x\in H^2(X;\BZ)$  with $x\equiv w_2(X) \hbox{ mod }2$
and $n_x$ is an integer associated with $x$.
For each $x$ we have an associated holomorphic line bundle
$L$ such that $x = - c_1(L^2) = -2 c_1(L)$. For each $L$ we have
the Seiberg-Witten monopole equation. The dimension of the
moduli space $\CM_{SW}$ of that monopoles
is given by
\eqn\swa{
dim_\BR \CM^x_{SW} = W_x =
\Fr{x\cdot x -(2\chi + 3\s)}{4}.
}
The amazing fact is that the moduli space is compact.
If $W_x = 0$ such that $x^2 = 2\chi + 3\s$, one simply counts
the number of the points with sign according to a suitable
orientation. The integer $n_x$ is that algebraic sum of the
points. If $W_x\neq 0$ and $W_x = 2 a$, one defines
\eqn\swb{
n_x = \int_{\CM^x_{SW}}\nu(\S)^a,
}
where $\nu : H_2(X;\BZ) \rightarrow H^2(\CM_{SW};\BZ)$ analogous
to the Donaldson $\mu$-map.
%We call such pairs $(x,n_x)$ a Seiberg-Witten
%invariant is $n_x$ if non-vanishing.
We would like to remark the followings.

i) The Seiberg-Witten invariants are metric independent for manifold
with $b_2^+ > 1$. Witten completely determined the invariants
for K\"{a}hler surface with $b_2^+ > 1$. For K\"{a}hler surface with
$p_g > 1$, all the higher dimensional Seiberg-Witten invariants
vanish (the simple condition). On the other hand, it is not known
if such a property still hold for manifold with $b_2^+ =1$.

ii) The Seiberg-Witten invariants vanish if a metric admits
positive scalar curvature. It is easy to see that the K\"{a}hler
surface with $p_g > 1$ does not admit a metric of positive
curvature except for the hyperK\"{a}hler case.  So there are at least
two Seiberg-Witten invariants including $(K_X, 1)$ and $(-K_X, -1)$.

iii) The precise relation between the Seiberg-Witten invariants
and the Donaldson invariants for manifold of simple type with
$b_2^+ \geq 3$ has been established by Witten \WittenE.
The Seiberg-Witten class $x$ corresponds to the basic class of
Kronheimer-Mrowka and $n_x$ to the coefficient associated
to the basic class \KM. The only other informations in the Donaldson
invariant are all homotopy invariants, such as the intersection form.

Now we come back to our original problem for a simply connected
K\"{a}hler surface with $p_g = 0$.
Similarly to the Donaldson invariants, the Seiberg-Witten
invariants $(x, n_x)$ is a topological invariants if $n_x$
does not change for a smooth path of metric joining two
generic metrics. It turns out $n_x$ jumps by $\pm 1$ if
there is a metric such that the monopole equation reduces
to the equation for the abelian instanton \WittenE\KMb\MorganB. This amounts
to
\eqn\swc{
\int_X x\wedge \o_g = 0.
}
Since $\o_g$ belongs to the positive cone, the above can happen if
and only if $x\cdot x < 0$, as we discussed before.
Conversely, $n_x$ is metric independent if $x\cdot x \geq 0$.

A necessary condition that $n_x $ can be non-vanishing
is the non-negativity of the dimension $W_x$,
\eqn\swd{
x\cdot x \geq 2\chi + 3\s.
}
Using $2\chi + 3\s = 9 -n$ for manifold of type $(1,n)$ with
$H^1(X;\BR) = 0$, one can conclude that all the
Seiberg-Witten invariants are metric independent for manifold
with $n\leq 9$ (or equivalently $b_2^- \leq 9$).

Now consider the manifold of type $(1, 9 + N)$ with $N > 0$.
Any possibly non-vanishing, at least for certain metric, Seiberg-Witten
class $x$ with $x \equiv w_2(X) \hbox{ mod } 2$ should satisfy
\eqn\swe{
x\cdot x \geq - N.
}
One can divide the set of all such $x$ into $\{x_{inv}\}\cup\{x^\pr\}$
where $x_{inv}\cdot x_{inv} \geq 0$. Clearly $n_{x_{inv}}$ is
metric independent. One can easily show that all the $n_{x_{inv}}$
are identically vanish. Recall that all $n_x$ vanish for
any $x$ with a metric which admits positive scalar curvature.
{}From $c_1(X)\wedge \o_g  =-K_X\wedge \o_g= \Fr{1}{2} R_g\o_g\wedge \o_g$
where $\o_g\wedge \o_g$ is positive definite, we have
\eqn\swf{
R_g < 0,\quad\hbox{for}\quad K_X\cdot [\o_g] > 0.
}
Assume that $X$ has a metric $g$ with negative scalar curvature,
i.e, $K_X\cdot [\o_g] > 0$. Since $K_X \cdot K_X = - N < 0$,
there is a period point $[\o_{g^\pr}]$ in the positive cone for some
metric $g^\pr$ such that $[\o_{g^\pr}] = K_X^{\perp}$, i.e.,
$K_X\cdot [\o_{g^\pr}] = 0$. Furthermore there exists a metric $g^\ppr$
such that $K_X\cdot [\o_{g^\ppr}] < 0$. Thus the positive cone
always has a chamber whose associated metrics admit positive scalar
curvature. It follows that all $n_x$ vanish at that chamber. Since
$n_{x_{inv}}$ is metric independent it should vanish identically.
Consequently, every possibly non-vanishing $x$ at least for certain
metric satisfies
\eqn\swg{
-N\leq x\cdot x \leq -1 \qquad\hbox{and}\qquad
x\equiv w_2(X)\hbox{ mod } 2.
}
Now we can conclude that
for manifold of type $(1, 9 + N)$  there are no Seiberg-Witten
class which are independent of the metric.\foot{This does not necessarily mean
that there are not genuine diffeomorphism invariants.}
Similarly to the Donaldson invariants, the Seiberg-Witten invariant
depends only on the chamber structure defined by $x$ satisfying
\swg\ and $x \equiv w_2(X)\hbox{ mod } 2$.

Consider one parameter family of $g_t$ joining two generic metrics $g_{-1}$
and
$g_1$ with $g_o$ such that $[\o_{g_0}] \in x^\perp$ and $g_t$ cross
the wall $x^\perp$ transversely. Witten showed that
\eqn\swh{
n_x(g_{-1}) = n_x(g_1) \pm 1.
}

\medskip\subsec{\bf The expectation values in TYM theory}

At present we do not know the precise relation between the Seiberg-Witten
invariants and the expectation values
$\left<\prod_{l=1}^{d-2r} \mu(\S_l)\,\Theta^r\right>_{\bf A}$. In any case,
$\left<\prod_{l=1}^{d-2r} \mu(\S_l)\,\Theta^r\right>_{\bf A}$ should be, in
general,
polynomials of degree $d$ in $H_2(X;\BZ)$ depending on $(x,n_x)$ and
$q_X$.  Since the values at the both two branches are well-defined,
we can confirm Witten's original claim that the path integral approach is
well-defined
whatever properties the moduli space of ASD connections has \WittenA.

If we consider manifold of type $(1,n)$ with $n\leq 9$, the variation
of the expectation value \qqe\ of TYM theory  comes only from
the branch {\bf B}. The relevant chamber structure in this case
is the set $\CC^k_X$ of the connected components  in the positive cone
after  removing  the system $\CW_k$ of walls defined by all the two-dimensional
integral classes $\zeta_i$ satisfying $\zeta_i\cdot\zeta_i = -k$.
The expectation values $\left<\prod_{l=1}^{d-2r} \mu(\S_l)\,\Theta^r\right>$
depend on metric only by the chamber structure $\CC^k_X$.
Let $C_+$ and $C_-$ be the two chambers  separated by the single wall $W_\zeta$
with the same condition as \baf.
{}From \qqe,
we have
\eqn\uaa{
\left<\prod_{l=1}^{d-2r} \mu(\S_l)\,\Theta^r\right>_{C_+}
-\left<\prod_{l=1}^{d-2r} \mu(\S_l)\,\Theta^r\right>_{C_-}
=  \Fr{2}{(2\pi)^{d-1}}2^{3r}  \prod_{l=1}^{d-2r}(\zeta\cdot \S_l).
}

For a manifold of type $(1,9+N)$ with $N>0$, the general  transition formula
can
be more complicated. We should consider the both chamber structures
for the branches ${\bf A}$ and ${\bf B}$.
Assuming that the metric crosses  a single wall $W_\zeta$ which does not
intersect with the walls $x^\perp$ defined by $x$ satisfying \swg,
the transition formula is given by \uaa.

\newsec{The Relations with the Donaldson Invariants}

In this section, we discuss some conjectural relations between the
expectation
values of TYM theory and the corresponding Donaldson invariants.
Clearly, the expectation value $\qqe$ does not coincide to the
corresponding Donaldson invariants
$\overline{q}_{k,X}(\S_1,\ldots,\S_{d-2r},(pt)^r)$.
Throughout this paper, we have tried to convince the (especially the
mathematician)
readers that the path integral approach to the gauge theoretic invariants
are well-defined whatever properties the moduli space of ASD connections
has.  It is clear, due to Witten, that the TYM theory correctly determines the
Donaldson
invariants at least for manifold with $b_2^+ \geq 3$.
For a manifold with $b_2^+ = 1$, the expectation values of
TYM theory in general have additional contributions essentially due to the
reducible
connections. We have shown that the expectation values are well-defined
and those additional contributions can be determined exactly.
In any case, the path integral approaches do not  refer to the compactification
procedure to get well-defined results.

Now the underlying reason for the discrepancy between the expectation
values of TYM theory and the Donaldson invariants is clear.
If we incorporated compactification of the moduli space, the path integral
will receive additional contributions from the reducible connections
with lower instanton numbers. Similarly to the expectation value of
the TYM theory, one can divide the Donaldson invariant
into the sum of two branches
\eqn\haa{
\overline{q}_{k,X}(\S_1,..,\S_{d-2r},(pt)^r)
 =\overline{q}_{k,X}(\S_1,..,\S_{d-2r},(pt)^r)_{\bf A}
+ \overline{q}_{k,X}(\S_1,..,\S_{d-2r},(pt)^r)_{\bf B}.
}
The equivalence between the TYM theory and Donaldson theory
for the manifolds with $b_2^+\geq 3$ implies that
\eqn\hab{
\overline{q}_{k,X}(\S_1,..,\S_{d-2r},(pt)^r)_{\bf A} =
\left<\prod_{l=1}^{d-2r} \mu(\S_l)\,\Theta^r\right>_{\bf A}.
}
Then the remaining part $\overline{q}_{k,X}(\S_1,..,\S_{d-2r},(pt)^r)_{\bf B}$
should be expressed as the sum of contributions of  reducible holomorphic
connections with the instanton numbers $1,2,....,k$. In particular,
the expression  $\left<\prod_{l=1}^{d-2r} \mu(\S_l)\,\Theta^r\right>_{\bf B}$
can be regarded as the contribution of the reducible holomorphic connections
with the instanton number $k$;
\eqn\hac{
\overline{q}_{k,X}(\S_1,..,\S_{d-2r},(pt)^r)_{\bf B} =
\Fr{1}{2} \sum_{\matrix{{}_{\zeta_i\cdot H < 0,} \cr  {}_{\zeta_i\cdot\zeta_i =
-k}\cr}}
 \Fr{1}{2^{3r}}   \prod_{l=1}^{d-2r}(\zeta_i\cdot \S_l) +\ldots,
}
where we normalized the expectation value by multiplying an universal factor
$$
(-1)^k\Fr{(2\pi)^{4k-4}}{2}.
$$

Throughout this section, we always assume that the path
of metric does not cross the walls responsible for the changes
of the Seiberg-Witten invariants.
We change the metric such that our ample class $H$  cross just one wall
$W_\zeta$,  defined by a certain divisor $\zeta$  satisfying
$\zeta \cdot H_+ <  0$ and $\zeta\cdot\zeta = -k$, and move
to another chamber $C_-$ with $\zeta\cdot H_- >  0$.
Provided that the path of metric does not cross another walls defined by
divisors $\zeta^\pr$ satisfying $\zeta^\pr\cdot \zeta^\pr = -1,...,-k+1$,
we immediately have the  transition formula
\eqn\cah{
\overline{\Gamma}^{k,r}_X (C_+)  - \overline{\Gamma}^{k,r}_X(C_-)
=  (-1)^{k}\Fr{1}{2^{3r}}\zeta^{d-2r}.
}
or
\eqn\cahh{
\overline{q}_{X,k,C_+}(\S^{d-2r},(pt)^r) -
\overline{q}_{X,k,C_-}(\S^{d-2r},(pt)^r)
= (-1)^{k}\Fr{1}{2^{3r}}(\zeta\cdot\S)^{d-2r}.
}
The transition formula \cah\ is a generalization of the formulas
obtained by Mong \Mong\ and Kotschick \KotsB\ for $r=0$  by Friedman and Qin
\FQB\
for the case $r=1$. It also agrees with the results of  Ellingsrud and
G\"{o}ttsche \EG for general values of $r$ up to certain normalization
difference.

\medskip\subsec{\bf The problem associated with the compactification}

Now the question is to determine the  contribution of the reducible
connections from the lower stratas in the compactified moduli space.
If we compactify  the moduli space, the path integral will get additional
contributions
from the reducible critical points whose associated divisors satisfying $-k + 1
\leq \zeta\cdot  \zeta < 0$.
Thus, we will simply view our result as the contribution
from the top strata.
To determine the extra contributions,
at least partially,  we will use  the recent result of Hu  and Li \HL
which shows that
\eqn\cba{
\overline{\CM}_k =
\bigcup_{\ell=0}^{k-1} \CM_{k-\ell}\times \hbox{Sym}^{\ell}(X),
}
if $k$ is sufficiently large (the condition that the K\"{a}hler metrics
on $X$ behave as generic metrics).

We consider a
$d$-dimensional subspace
$\CN_{k-\ell} = N_{k-\ell} \times  \hbox{Sym}^{\ell}(X) \subset
\overline\CM_k$
such that $\CM_{k-\ell} \subset N_{k-\ell}$,
$\hbox{dim}_\BC(N_\ell) =d-2\ell$.
We also assume that $N_{k-\ell}\backslash \CM_{k-\ell}$ does not
contains any reducible connections, for arbitrary   metric, other than
$\CM_{k-\ell}\backslash \CM^*_{k-\ell}$.  We
consider $\bigcup_{\ell=0}^{k}\CN_{\ell}\subset \overline{\CM}_k$.
The restriction of $\overline\mu(\S)$ to
$N_{k-\ell} \times  \hbox{Sym}^{\ell}(X)$ will be in the form
\eqn\cbb{
\overline\mu(\S)_\ell = \mu(\S)_\ell + 2 H^\pr,
}
where $H^\pr \in H^2(\hbox{Sym}^{\ell}(X))$.
We can consider the part of the Donaldson invariants contributed from
$\CN_{k-\ell}$,
\eqn\cbc{
\eqalign{
\left< \overline\mu(\S)^d,  \CN_{k-\ell}\right>
&=2^{2\ell} \Fr{d!}{(d-2\ell)!(2\ell)! }\left< \mu(\S)_\ell^{d-2\ell},
N_{k-\ell}\right>
    \left<(H^\pr)^{2\ell}, \left[\hbox{Sym}^{\ell}(X)\right]\right>\cr
&= 2^{\ell} \Fr{d!}{(d-2\ell)!\ell! }\left< \mu(\S)_\ell^{d-2\ell},
N_{k-\ell}\right>
     q^\ell, \cr
}
}
where we have used
\eqn\cbd{
\left<(H^\pr)^{2\ell}, \left[\hbox{Sym}^{\ell}(X)\right]\right>
= \Fr{(2\ell)!}{2^\ell\ell!} q^\ell, \hbox{ where } q \in
\hbox{Sym}^2(H^2(X;\BZ)).
}
Now we can use our previous result to determine
$\left< \mu(H)_\ell^{d-2\ell}, N_{k-\ell}\right>_{\bf B}$, which gives
\eqn\cbe{
\Fr{1}{2}
\sum_{\zeta_i\cdot H < 0}
(-1)^{k-\ell}\left(\zeta_i\cdot \S\right)^{d-2\ell}.
}
where the summation runs over every divisor satisfying
$\zeta_i \cdot \zeta_i = -k+\ell$.
By summing up, we can write
\eqn\cbf{
\overline{q}_{X,k,C(X)}(\S^{d})_{\bf B}
=\Fr{1}{2}\sum_{\ell=0}^{k}
(-1)^{k-\ell}2^{\ell}\!\!\!\!\!\!\!\!\sum_{\matrix{{}_{\zeta_i\cdot H < 0,}
\cr
{}_{\zeta_i\cdot\zeta_i = -k +\ell}\cr}} \Fr{d!}{(d-2\ell)!\,\ell!}
\left(\zeta_i\cdot \S\right)^{d-2\ell} q^\ell + \ldots
}
In other words
\eqn\cbg{
\overline \G^{ k}_X (C)_{\bf B} =
\Fr{1}{2}\sum_{\ell=0}^{k}
(-1)^{k-\ell}2^{\ell}\!\!\!\!\!\!\!\!\sum_{\matrix{{}_{\zeta_i\cdot H < 0,}
\cr
{}_{\zeta_i\cdot\zeta_i = -k +\ell}\cr}} \Fr{d!}{(d-2\ell)!\,\ell!}
(\zeta_i)^{d-2\ell} q^\ell + \ldots
}

Let a smooth path of K\"{a}hler metric meet only one wall $W_\zeta$
defined by $\zeta$ such that
\eqn\cbh{
\zeta\cdot\zeta = -k + \ell_\zeta, \qquad
\zeta\cdot H_+ < 0 < \zeta\cdot H_-.
}
We have
\eqn\cbi{
\overline\G^{k}_X (C_+) -\overline\G^{ k}_X (C_-)  =
(-1)^{k-\ell_\zeta}2^{\ell_\zeta}
\Fr{d!}{(d-2\ell_\zeta)!\,\ell_\zeta!}
(\zeta)^{d-2\ell_\zeta} q^{\ell_\zeta} + \ldots
}

Now we can easily generalize the result to the polynomials including  the
four-dimensional
class
\eqn\cbj{\eqalign{
\overline{q}_{X,k}&(\S^{d-2r}(pt)^r )_{\bf B}\cr
&=
\Fr{1}{2}\sum_{\ell=0}^{k}
(-1)^{k -\ell}
2^{\ell-3r}\Fr{ (d-2r)!\,r!}{(d-2\ell-2r)!\,\ell!}
\!\!\!\!\!\!\!\!\sum_{\matrix{{}_{\zeta_i\cdot H < 0,} \cr
{}_{\zeta_i\cdot\zeta_i = -k +\ell}\cr}}\!\!\!\!\!\!
\left(\zeta_i\cdot \S\right)^{d-2\ell-2r}q^\ell
+\cdots,\cr
}
}
where $r = 0,1,..,[d/2]-\ell$.
In other words,
\eqn\cbk{
\overline \G^{k, r}_X (C)_{\bf B} =
\Fr{1}{2}\sum_{\ell=0}^{k}
(-1)^{k -\ell}
2^{\ell-3r}\Fr{ (d-2r)!\,r!}{(d-2\ell-2r)!\,\ell!}
\!\!\!\!\!\!\!\!\sum_{\matrix{{}_{\zeta_i\cdot H < 0,} \cr
{}_{\zeta_i\cdot\zeta_i = -k +\ell}\cr}}\!\!\!\!\!\!
(\zeta_i)^{d-2\ell-2r}q^\ell
+\cdots.
}
Thus the transition formula is given by
\eqn\cbl{
\overline\G^{ k,r}_X (C_+) -\overline\G^{ k,r}_X (C_-)
=(-1)^{k -\ell_\zeta}
2^{\ell_\zeta-3r}\Fr{ (d-2r)!\,r!}{(d-2\ell_\zeta-2r)!\,\ell_\zeta!}
(\zeta)^{d-2\ell_\zeta-2r}q^{\ell_\zeta}
+\cdots.
}
This  formula has been obtained by
Kotschick and Morgan \KoM\ for $r=0$, by Friedman and Qin \FQB\
for $r=1$ and by Ellingsrud and G\"{o}ttsche \EG\ for general $r$.
The paper  \FQB\ has some explicit results beyond the
leading term. In the paper \EG, up to the leading $3$-terms
were calculated for general $r$.

\medskip\subsec{\bf The variation of the moduli space}

It is well-known that the image of moment map is a convex cone in a positive
Weyl chamber of the Lie algebra and  the critical values of the moment map
define a system of walls in the convex cone \Atiyah.  It is also known that
the symplectic quotients undergo a specific birational transformations
closely related to the variation of GIT quotients as the values of moment
map cross the wall\GS\ThB\DoH. The walls are determined by the critical
points of the moment map. Of course, all these are rigorous for finite
dimensional compact manifold with compact group actions.

Formally speaking, our problem is an infinite dimensional analogue
of the variation of the symplectic quotients. The moduli space of
ASD connection can be identified with the symplectic quotients
$\eufm{m}(0)^{-1}/\CG$.  An interesting fact is that
the chamber structures in the positive cone of the manifold and
those in the convex cone  sitting on image of the moment map
are determined by the same data.
Our  results   imply that the two chamber structures are  isomorphic!
Naively speaking, this suggests that
the moduli space of ASD connections undergoes  certain
birational transformations if the metric cross that walls.
This picture also coincides with our general strategy for deriving
the transition formula. Losing a higher critical point $\zeta$
and getting another  higher critical point $-\zeta$ is analogous to the
blown-down
and successive blown-up.

In the algebro-geometrical method, one constructs
the moduli space of $H$-stable bundles over algebraic surface
and studies the variation of the moduli space under the changes of
the polarizations \Li\Morgan. One of the advantage of the
algebro-geometrical approach is that it has a natural way of
the compactification of the moduli space.  Since the moduli space of
$H$-semi-stable bundles contains  the moduli space of the $H$-stable bundles
as a Zariski open subset,  it gives a natural compactification.
It is well-known that the diffeomorphism class of the moduli space depends on
the chamber structure  in the ample (K\"{a}hler) cone.
Recently several papers on the variation of the
moduli space of $H$-semi-stable bundle under the changes of the polarization
of the ample class appeared \EG\MW\FQB.
They show that the moduli space undergoes specific birational transformation
similar to the variation of geometrical invariant theory (GIT)
quotients studied by Thaddeus \ThA\ThB\ and Dolgachev-Hu \DoH.
The  variation of GIT quotients was also studied independently by Witten in the
context of two-dimensional supersymmetric theory and the quantum cohomology
rings\WittenF.  We note that Witten's picture is quite similar to our approach.

It is not yet clear if a natural compactification of the moduli
space can be obtained using some path integral methods. However,
our result is sufficient to predict the general form of the
Donaldson invariants as \haa\hab\hac.
If we consider when the variation of the Donaldson invariants
get contribution only from $\overline q_{X,k,C(X)}(\S^{d-2r}(pt)^r)_{\bf B}$,
the transition formula of the Donaldson invariants
is sufficient to determine  $\overline q_{X,k,C(X)}(\S^{d-2r}(pt)^r)_{\bf B}$
as one can see from the relation between \cbl\ to \cbk\ and \cbj.

\listrefs
\bye